\documentclass[reprint,twocolumn,secnumarabic,amssymb, aps, prx]{revtex4-2}

\usepackage{graphicx}
\usepackage{gensymb}
\usepackage{amsmath}
\usepackage{braket}
\usepackage{lipsum}
\usepackage{xcolor}
\usepackage{float}
\usepackage{wrapfig}

\newcommand{\appropto}{\mathrel{\vcenter{
  \offinterlineskip\halign{\hfil$##$\cr
    \propto\cr\noalign{\kern2pt}\sim\cr\noalign{\kern-2pt}}}}}

\makeatletter
\newcommand*{\newbibstartnumber}[1]{%
  \apptocmd{\thebibliography}{%
    \global\c@NAT@ctr #1\relax
    \addtocounter{NAT@ctr}{-1}%
  }{}{}%
}
\makeatother

\usepackage{hyperref}
\hypersetup{
    pdfstartview={FitH},    
    colorlinks=true,       
    linkcolor=blue,          
    citecolor=blue,        
    filecolor=magenta,      
    urlcolor=blue           
}

\begin{document}
\newcommand\numberthis{\addtocounter{equation}{1,2}\tag{\theequation}}
\title{Quantum coarsening and collective dynamics on a programmable simulator}
\author{
Tom~Manovitz$^{1,*}$, Sophie~H.~Li$^{1,*}$, Sepehr~Ebadi$^{1,*,\ddagger}$, Rhine~Samajdar$^{2,3}$, Alexandra~A.~Geim$^{1}$, Simon~J.~Evered$^{1}$, Dolev~Bluvstein$^{1}$, Hengyun~Zhou$^{1,4}$, Nazli~Ugur~Koyluoglu$^{1,5}$, Johannes~Feldmeier$^{1}$, Pavel~E.~Dolgirev$^{1}$, Nishad~Maskara$^{1}$, Marcin~Kalinowski$^{1}$, Subir~Sachdev$^{1}$, David~A.~Huse$^{2}$, Markus~Greiner$^{1}$, Vladan~Vuleti\'{c}$^{6}$, and Mikhail~D.~Lukin$^{1,\dagger}$}

\affiliation{$^1$Department~of~Physics,~Harvard~University,~Cambridge,~MA~02138,~USA \quad \quad\\
$^2$Department of Physics, Princeton University, Princeton, NJ 08544, USA\\
$^3$Princeton Center for Theoretical Science, Princeton University, Princeton, NJ 08544, USA\\
$^4$QuEra Computing Inc., Boston, MA 02135, USA\\
$^5$Harvard Quantum Initiative, Harvard University, Cambridge, MA 02138, USA\\
$^6$Department~of~Physics~and~Research~Laboratory~of~Electronics,~Massachusetts~Institute~of~Technology,~Cambridge,~MA~02139,~USA\\
$^\ddagger$Current address: Department~of~Physics,~Massachusetts~Institute~of~Technology,~Cambridge,~MA~02139,~USA\\
$^*$These~authors~contributed~equally~to~this~work~$^\dagger$Corresponding~author;~E-mail:~lukin@physics.harvard.edu
}

\begin{abstract}

Understanding the collective quantum dynamics of nonequilibrium many-body systems is an outstanding challenge in quantum science. In particular, dynamics driven by quantum fluctuations are important for the formation of exotic quantum phases of matter \cite{altman2023quantum}, fundamental high-energy processes \cite{bauer2023highenergy}, quantum metrology \cite{degen2017sensing, li2023scrambling}, and quantum algorithms \cite{ebadi2022quantum}. Here, we use a programmable quantum simulator based on Rydberg atom arrays to experimentally study collective dynamics across a (2+1)D Ising quantum phase transition. 
After crossing the quantum critical point, we observe a gradual growth of correlations through coarsening of antiferromagnetically ordered domains~\cite{Samajdar2024}. By deterministically preparing and following the evolution of ordered domains, we show that the coarsening is driven by the curvature of domain boundaries, and find that the dynamics accelerate 
with proximity to the quantum critical point. 
We quantitatively explore these phenomena and  
further observe long-lived oscillations of the order parameter, corresponding to an amplitude (`Higgs') mode \cite{pekker2015amplitude}. These observations offer a unique viewpoint into emergent collective dynamics in strongly correlated quantum systems and nonequilibrium quantum processes.

\end{abstract}

\maketitle

Quantum phase transitions (QPTs) are transformations between states of matter that are driven by quantum fluctuations \cite{sachdev2011quantum}. Analogously to thermal fluctuations in classical phase transitions, quantum fluctuations play a dominant role in 
the emergence of order in quantum systems. 
While classical dynamics near thermal critical points have been studied extensively over the past several decades, only recently have quantum  dynamics across QPTs become experimentally accessible, due to the advent of quantum simulators \cite{bakr2010probing,keesling2019quantum,ebadi2021quantum,scholl2021quantum} and ultrafast spectroscopic methods in solid-state systems \cite{ruegg2008quantum,pekker2015amplitude,jain2017higgs,shimano2020higgs}. Their universal properties have been studied in systems of varied dimensionality using the 
quantum Kibble-Zurek mechanism (KZM) \cite{pyka2013topological,keesling2019quantum,ebadi2021quantum}. 
The KZM stipulates that a quantum system's dynamics and correlations ``freeze''
in the vicinity of a QPT when the system can no longer respond adiabatically to dynamical changes. 
However, in many instances, other mechanisms of correlation growth beyond KZM can dominate ordering\cite{biroli2010kibble,roychowdhury2021dynamics,king2022coherent,schmitt2022quantum,zeng2023universal}.  
In particular, when an unordered system passes through a continuous phase transition into a symmetry-broken phase, a progressive growth of long-range order, known as coarsening, is expected.
These ordering dynamics are predicted to exhibit universality, manifested as self-similarity in the growth of correlations \cite{lifshitz1962kinetics,hohenberg1977critical,Bray1994}.
Such phenomena are well understood in classical systems \cite{Bray1994}, and have been experimentally explored in Bose gases in the mean-field regime over the past two decades \cite{sadler2006spontaneous,PhysRevX.8.021070,prufer2018observation,erne2018universal,goo2022universal,gazo2023universal,huh2024universality}. However, effects of quantum fluctuations in coarsening dynamics, particularly near QPTs, have only recently emerged as a subject of  theoretical \cite{chandran2012kibble,chandran2013equilibration,maraga2015aging,gagel2015universal,Samajdar2024} and experimental \cite{andersen2024thermalization} investigation.

We use a programmable quantum simulator based on Rydberg atom arrays to investigate the collective out-of-equilibrium dynamics associated with the growth of order following an Ising QPT. We observe key features of beyond-mean-field quantum coarsening processes arising from quantum fluctuations: the curvature-driven dynamics of domain walls and their acceleration when approaching the critical point.
We further explore self-similarity and universality in the ordering process. 
Additionally, we observe long-lived coherent oscillations of the correlation length and the order parameter on both sides of the QPT.  In the ordered phase, these oscillations are the analog of a ``Higgs'' mode 
\cite{shimano2020higgs,ruegg2008quantum,endres2012higgs}. Our observations are  consistent with theoretical predictions \cite{sachdev2011quantum}, extending these studies in a regime that is difficult to simulate classically. 

\begin{figure*}
\includegraphics[width=2\columnwidth]{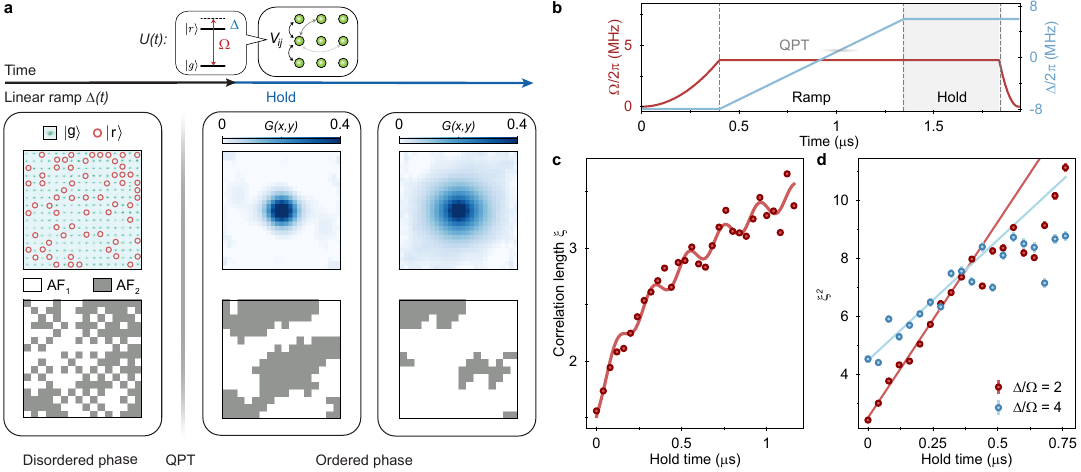} 
\caption{\textbf{Ordering via coarsening.} 
\textbf{a, } Atoms arranged in a square lattice are prepared in the ground state $\ket{g}$ and driven from the disordered phase, shown here as a fluorescence image (atoms in $\ket{r}$ are detected as loss and are indicated with red circles), 
into an antiferromagnetic ordered phase, across a quantum phase transition. The subsequent coarsening dynamics lead to the progressive development of long-range order (shown here using the real-space correlation function $G(x,y)$ and experimental snapshots). The snapshots show the local single-spin-flip-corrected staggered magnetization, with white (dark grey) shading indicating AF\textsubscript{1} (AF\textsubscript{2}) orders (see Methods).
\textbf{b, } The detuning $\Delta(t)$ is swept linearly, with a fixed Rabi frequency $\Omega$, to a final value $\Delta$ in the ordered phase, where it is held constant for the duration of the hold time. 
\textbf{c, } The correlation length $\xi$ grows over the course of the hold time. The solid line is a fit to $(c_0+c_1 t)^{\alpha} + c\cos{(\omega t+\phi)}$, representing sinusoidal oscillations on top of a power-law growth. Errorbars are smaller than markers.
\textbf{d, } The early-time growth of the squared correlation length is consistent with a linear increase with time, as expected for coarsening dynamics with a nonconserved order parameter. The dynamics are faster at lower $\Delta/\Omega$, closer to the critical point, see also ED Fig. \ref{fig:corr_lengths}c and \ref{fig:corr_lengths}d.
For the data shown here, $\Omega/2\pi = 3.8$\,MHz. Errors bars are given as the standard deviation, unless noted otherwise.} 
\label{fig:correlations}

\end{figure*}

Our experiments are performed using a two-dimensional programmable atom array, previously described in Ref.~\cite{ebadi2021quantum}. The measurements are conducted on a $16\times16$ square lattice of  $^{87}\textrm{Rb}$ atoms trapped in an array of optical tweezers generated by a spatial light modulator (SLM). Atoms are initialized in the electronic ground state $\ket{g}$ and are coupled to the high-lying electronic Rydberg state $\ket{r}$ through a two-photon excitation with time-dependent Rabi frequency $\Omega (t)$ and global detuning $\Delta (t)$. As a key upgrade, we introduce a second SLM for generating locally controlled light shifts, allowing for programmable site-dependent detunings $\delta_i (t) = \alpha_i \delta (t)$ (see Methods and Extended Data (ED) Fig.~\ref{fig:localcontrol_supp}).
The atoms in the $\ket{r}$ state interact strongly  via a van der Waals potential, giving rise to the following Hamiltonian governing the system's dynamics:
\begin{equation}
    \frac{H}{\hbar} = \frac{\Omega(t)}{2}\sum_i X_i -  \sum_i n_i (\Delta(t) +\delta_i(t)) + \sum_{i<j} V_{ij}n_i n_j.
\label{eq:ryd_ham}
\end{equation}
Here, $n_i$\,$\equiv$\,$\ket{r_i}\bra{r_i}$ denotes the Rydberg occupation at site $i$, $X_i \equiv \ket{g_i}\bra{r_i}+\ket{r_i}\bra{g_i}$ describes the laser-induced coupling between the states at that site, and $V_{ij} \equiv V_0/\lvert \boldsymbol{r}_i - \boldsymbol{r}_j \rvert^6$ is the van der Waals interaction. The Rydberg interactions prevent simultaneous excitation of two atoms if they lie within a blockade radius ($R_b \equiv (V_0/\Omega)^{1/6}$) of each other. We maintain a lattice spacing $a$ such that $R_b/a\approx1.1$, and only nearest-neighboring sites fall within the blockade radius. For large positive values of $\Delta/\Omega$, this configuration leads to a $\mathbb{Z}_2$-symmetry-broken checkerboard phase with two antiferromagnetically ordered ground states \cite{PhysRevLett.124.103601}, labeled $\ket{\textrm{AF}_1}$ and $\ket{\textrm{AF}_2}$. The disordered and ordered phases are separated by a QPT belonging to the (2+1)D Ising universality class, which occurs at $\Delta/\Omega\approx 1.1$ \cite{PhysRevLett.124.103601,Kalinowski2022,ebadi2021quantum}.  The order parameter diagnosing this transition is the staggered magnetization: $m_s=\sum_{x,y}\tilde{Z}_{x,y}=\sum_{x,y}(-1)^{x+y} Z_{x,y}$, where $(x,y)$ denotes the two-dimensional coordinates of an atom, and $Z_{x,y} \equiv \ket{r_{x,y}}\bra{r_{x,y}}-\ket{g_{x,y}}\bra{g_{x,y}}$.

\section*{Ordering dynamics}
\label{sec:ordering}

We first study the nonequilibrium dynamics of the atom array after crossing the QPT into the ordered phase. Our protocol is illustrated in Fig.~\ref{fig:correlations}a,b. The state-preparation stage is similar to that used in Refs.~\cite{ebadi2021quantum, semeghini2021probing}. First, all atoms are initialized in $\ket{g}$, which is the ground state for $\Delta/\Omega\ll 0$. While keeping $\Delta$ negative, $\Omega$ is ramped up to its final value, remaining constant until the end of the protocol ($\delta_i$ is held at 0 for this measurement).  Then, $\Delta(t)$ is swept from negative to positive values, through the quantum critical point ($\Delta/\Omega \approx 1.1\,$\cite{ebadi2021quantum}) and into the ordered phase. The non-adiabatic sweep realizes a quench that injects energy into the system and seeds the ensuing nonequilibrium evolution. We use a linear sweep profile for all measurements; for a discussion of the sweep rate, see Methods. The sweep is halted at various endpoints within the ordered phase.  Subsequently, $\Delta$ and $\Omega$ are held constant for a given hold time, and finally $\Omega$ is ramped down followed by a projective readout of the atomic states.

During the hold time, we probe the dynamics of the correlation length, as shown in Fig.~\ref{fig:correlations}c.  
To quantify the growth of correlations, we evaluate the two-point correlation function $G(\textbf{r},t)$ and the radially averaged structure factor $S(k,t)$, from which we extract a correlation length ($\xi$) (see Methods and ED Fig.~\ref{fig:corr_lengths}a,b).
In contrast to the Kibble-Zurek prediction, the correlation length grows significantly with hold time (Fig.~\ref{fig:correlations}c), indicating the gradual establishment of long-range order. 
Up to a hold time of $\sim 0.4$--$0.5$\,$\mu s$,  we  observe that the dynamics are consistent with a linear growth of $\xi^2$ with time (Fig.~\ref{fig:correlations}d), as expected for coarsening \cite{Bray1994, Samajdar2024}. Importantly,  the rate of growth increases with proximity to the QPT, an observation further explored in ED Fig.~\ref{fig:corr_lengths}c,d.
Motivated by the theoretical expectation of universality of these dynamics in the thermodynamic limit~\cite{Bray1994,barenblatt1996scaling,gazo2023universal}, we also
study the structure factor
as a function of a scaling variable $k \,\xi(t)$. We find that the data collapse onto a single functional form  $S (k,t) \sim b(\xi(t))\xi^2 (t) f (k \,\xi(t))$ for some scaling function $f$ and an amplitude $b(\xi)$, which is suggestive of  self-similarity
(see ED Fig.~\ref{fig:collapse} and Methods). 
In addition to the ordering, we also observe long-lived oscillations of the correlation length (Fig.~\ref{fig:correlations}c). In what follows, the universal aspects and origin of these oscillations are explored in detail.

\begin{figure}
\centering
\includegraphics[width=1\columnwidth]{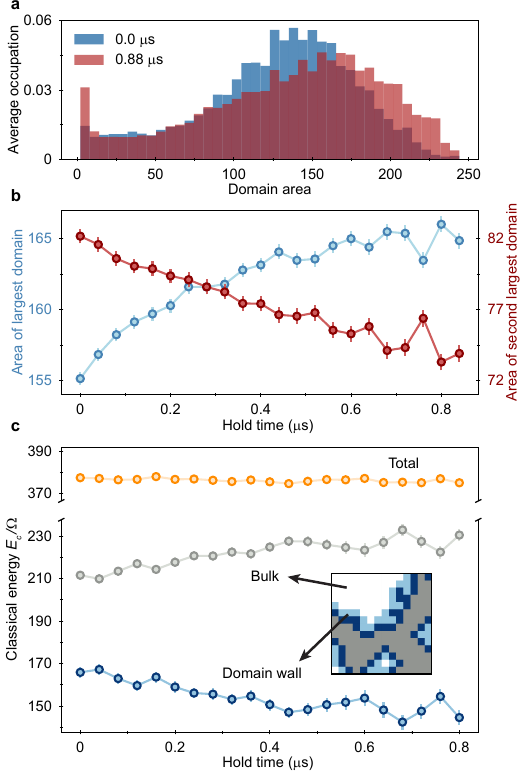} 
\caption{\textbf{Domains and energy transfer.}
\textbf{a,} Probability distribution of an atom belonging to an ordered domain of a certain size. At late hold times, atoms are more likely to participate in large domains and very small domains. \textbf{b, } The area of the largest domain (given in number of atoms) grows, while that of the second-largest domain decreases as the system is held in the ordered phase. \textbf{c, } Classical energy of each snapshot during the hold time, calculated using Eq.~\eqref{eq:Classical_Hamiltonian}. The total classical energy of the system is conserved while the bulk (domain wall) energy increases (decreases). Inset: separation of domains into bulk and domain walls for a single snapshot. The domain walls are identified as regions of the array where neither AF ordering is observed (light and dark blue). White and grey  indicate AF\textsubscript{1} and AF\textsubscript{2} orderings respectively. The data presented here are for $\Delta/\Omega = 3.0$, with $\Omega/2\pi=3.8$ MHz for \textbf{a,b} and $\Omega/2\pi=6.0$ MHz for \textbf{c}.
}
\label{fig:global coarsening}
\end{figure}

In order to gain further insight into the system's dynamics, we use single-site-resolved detection to identify the domains in each individual snapshot. We measure the probability that a given atom will appear as part of a domain of area $A_d$, and find that over time, increasingly larger domains are formed at the expense of their smaller counterparts (Fig.~\ref{fig:global coarsening}a). This is manifested in the growth of the mean area of the largest domain, concurrent with the shrinking of the second-largest one (Fig.~\ref{fig:global coarsening}b).
Due to energy conservation, the appearance of progressively larger domains has to be offset by the proliferation of very small domains and single-site spin flips, as apparent in Fig.~\ref{fig:global coarsening}a. 
We quantify this flow of energy by measuring the spatial distribution of the classical energy, as defined by the diagonal contribution to Eq.~\eqref{eq:ryd_ham} up to a constant shift:
\begin{equation}
    H_{\mathrm{cl}}= -\Delta\sum_i (n_i-1) + \sum_{i<j}V_{ij}n_in_j.
\label{eq:Classical_Hamiltonian}
\end{equation}
For every snapshot, we identify the domain walls and the bulk, and accordingly determine the contribution of each towards $\braket{H_{\mathrm{cl}}}$ (see Methods and ED Fig.~\ref{fig:local_analysis}a,b). While the classical energy is indeed conserved over time, it is redistributed from the domain walls into the bulk (Fig.~\ref{fig:global coarsening}c). This is consistent with a picture of coarsening which is driven by the surface tension, and elimination of domain walls \cite{Bray1994}.

\begin{figure*}
\centering
\includegraphics[width=2\columnwidth]{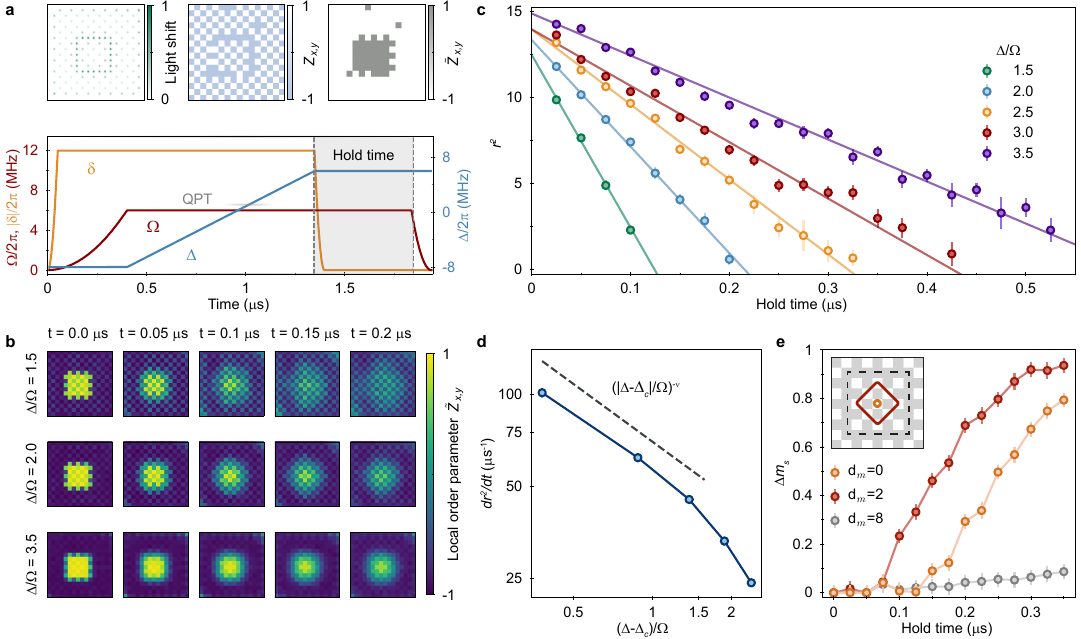} 
\caption{\textbf{Dynamics of seeded domains.} 
\textbf{a,} A centered domain of AF\textsubscript{2} order surrounded by a backdrop of AF\textsubscript{1} order is created by ``pinning" each ground-state atom in the target ordering using a local light shift $\delta_i$, while sweeping $\Delta (t)$.
Upper left: target amplitude of the local detuning pattern; $|\delta_i|$ is inversely proportional to the number of neighboring Rydberg atoms surrounding each ground-state atom. Center: Rydberg population of a single shot immediately after the local detuning is quenched off. Right: local staggered magnetization, demarcating the prepared domains. \textbf{b, } Evolution of the average local staggered magnetization with time, as the prepared state evolves under different values of $\Delta/\Omega$.  \textbf{c, } The squared radius of the central domain, $r^2$, decreases linearly with time (see also ED Fig.~\ref{fig:local_analysis}c). \textbf{d, } The rate of change of the area,  $r^2$, increases with proximity to the critical point, $(\Delta - \Delta_c)/\Omega$. The dashed line is a guide to the eye for the theoretically expected scaling relationship. \textbf{e, } Change in the radially averaged local staggered magnetization with Manhattan distances $d_m = 0, 2, 8$ from the center of the injected square domain. The order initially changes near the domain wall and begins to change in the center of the domain only at later times. Atoms in the bulk of the dominant ordering far from the domain wall, in gray, remain close to their initially prepared states. The dynamics are plotted for $\Delta/\Omega = 2.5$ and $\Omega/2\pi=6.0$ MHz.}

\label{fig3:square defect}
\end{figure*}

\section*{Domain wall dynamics}
\label{sec:domain}

In order to study the real-time dynamics of domains and domain walls, we deterministically prepare specific configurations of domain walls using programmable, locally controlled light shifts. Our protocol is described in Fig.~\ref{fig3:square defect}a: we apply site-dependent negative local detunings $\delta_{i}<0$, with amplitudes $|\delta| \sim 4\Omega$ (see Methods), on the chosen atoms prior to ramping up $\Omega $, and then continue the state-preparation protocol as previously described. The local detuning strongly biases the chosen atoms to $\ket{g}$, and consequently, locally favors either an $\ket{\textrm{AF}_1}$ or an $\ket{\textrm{AF}_2}$ configuration. After the sweep is completed, and before the hold time begins, the local detuning is quenched off and the state is allowed to evolve freely. 

We start by preparing a small square domain of one AF order within the bulk of the other (Fig.~\ref{fig3:square defect}b) \cite{pavevsic2024constrained}. Upon removal of the local detunings, we observe that the area of the injected domain shrinks linearly with time (Fig.~\ref{fig3:square defect}c and Supplementary Video 1). This observation is in agreement with coarsening dynamics for nonconserved fields, where surface tension due to the energy cost of domain walls generates curvature-driven dynamics \cite{lifshitz1962kinetics,Bray1994}. In such a scenario, the local velocity of a domain wall is proportional to its local curvature $1/R$ (where $R$ is the local radius of curvature): $\partial_t R \propto -1/R$, and therefore, $\partial_t R^2 = -v_a$, where $v_a$ is some positive time-independent constant. 
Strikingly, we find that $v_a$ increases as one approaches the quantum critical point. This behaviour is unique for coarsening in the vicinity of a \textit{quantum} phase transition \cite{Samajdar2024}; in contrast, for a classical Ising transition, the dynamics should be slower near the thermal phase boundary than when deep in the ordered phase \cite{humayun1991non}; 
we also observe indications of this speedup in the global sweeps (see Fig.~\ref{fig:correlations}d and Methods). We examine the dependence of $v_a$ on the distance to the quantum critical point, $\Delta-\Delta_c$  (Fig.~\ref{fig3:square defect}d). Near the QPT, we find that $v_a$ is approximately consistent with a scaling $\propto(\Delta-\Delta_c)^{-\nu}$(where $\nu\approx 0.629$ is the correlation length exponent of the (2+1)D Ising QPT \cite{PhysRevLett.124.103601}). We note that the speedup observed here is not caused by the KZM, as it only depends on proximity to the QPT, consistent with theoretical predictions \cite{Samajdar2024}. In Fig.~\ref{fig3:square defect}e, we also analyze the evolution of several concentric spatial layers of the system, and observe the outer layers of the central domain morphing earlier with the dynamics moving progressively inwards. This supports a picture of coarsening in which the dynamics are indeed driven by the shrinking of domain walls, as opposed to being generated within the bulk of the domains. 

\begin{figure}
\centering
\includegraphics[width=1\columnwidth]{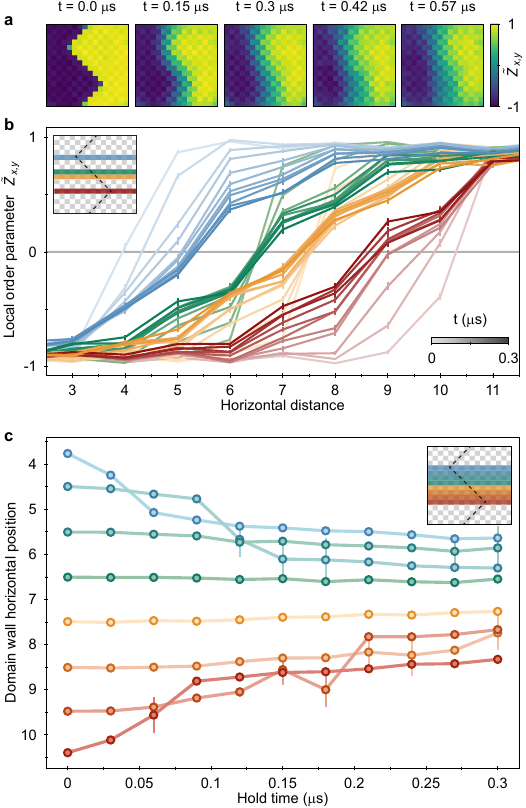} 
\caption{\textbf{Curvature-driven dynamics.} \textbf{a, } We create two separate AF orderings with a zigzag domain wall between them. During the hold time, the domain wall straightens into a vertical line.
\textbf{b, } Horizontal cuts of the staggered magnetization at points where the domain wall is locally curved (red, blue) versus straight (green, yellow). At points of high local curvature (red and blue) the domain wall moves towards the center of the domain, while at points of low local curvature (green and yellow) the domain wall remains stationary. \textbf{c, } The domain wall's horizontal position for each row (as indicated by the color) as a function of the hold time. In rows where the domain is locally curved, the domain wall moves towards the center. All data shown are with $\Omega/2\pi=6.0$ MHz; \textbf{b} is measured at $\Delta/\Omega=3.0$ and \textbf{a,c} at $\Delta/\Omega=2.5$.
}
\label{fig4:zigzag}
\end{figure}

To further explore the curvature-driven nature of the coarsening dynamics, we prepare an initial state with a zigzag domain wall. Over time, the domain wall straightens into a vertical line separating the two orders (Fig.~\ref{fig4:zigzag}a and Supplementary Video 2), while  the motion of the domain wall is related to its local curvature~\cite{Samajdar2024}, as shown in Figs.~\ref{fig4:zigzag}b,c and ED Fig.~\ref{fig:local_numerics} (see Methods).

\begin{figure*}
\centering
\includegraphics[width=2\columnwidth]{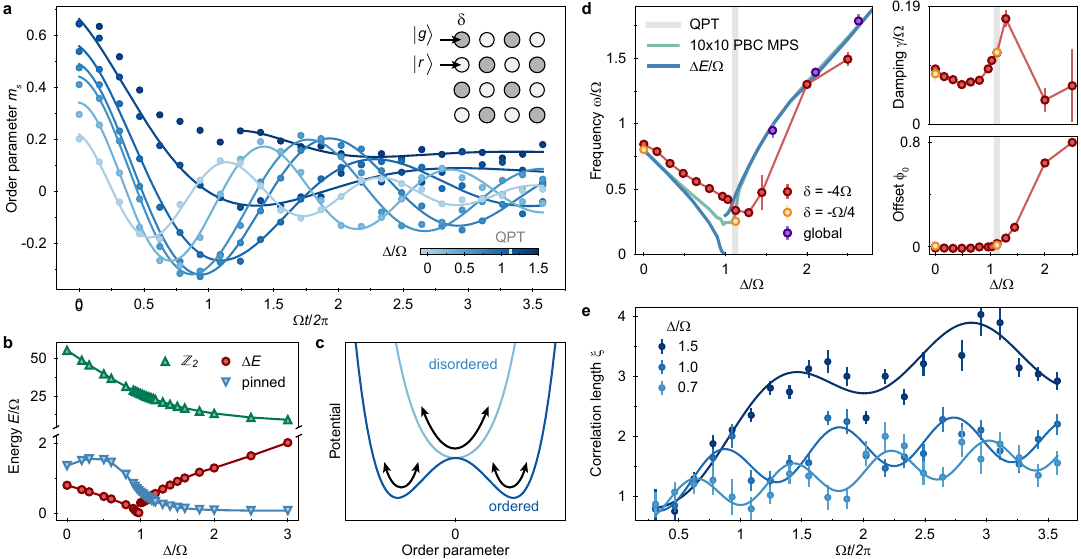} 
\caption{\textbf{Higgs mode oscillations.} \textbf{a,} Long-lived oscillations of the staggered magnetization observed when one sublattice is initially pinned and then released. 
\textbf{b,} Numerical calculation of the energy difference, in a $10\times10$ system with periodic boundary conditions, of the initial pinned state (blue) and the $\mathbb{Z}_2$ state (green) relative to the ground state, as well as the gap to the first excited state (red). The pinned state is much lower in energy than the $\mathbb{Z}_2$ state at finite detuning.  \textbf{c, } Schematic diagram of the effective potential for the amplitude mode. 
\textbf{d, } Oscillation parameters extracted from fitting $\phi(t) \approx \phi_0+A \cos{(\omega t+\theta_0)}e^{-\gamma t}$ to data shown in a. Left: Measured oscillation frequencies (red points), plotted along with numerically determined values and calculated ground-state energy gaps (both for a $10 \times10$ lattice). 
A lower local detuning $|\delta|$ leads to oscillations of lower amplitude and frequency (orange). Oscillation frequencies of the magnetization in the global sweeps, as shown in Extended Data Fig.~\ref{fig:higgs_global}a, are indicated in purple. Upper right: the measured oscillations decay more rapidly as the phase transition is approached. Lower right: in the disordered phase, oscillations occur around $m_s = 0$, while in the ordered phase, a finite offset emerges. \textbf{e, } Oscillations of the correlation length. For final detunings close to the quantum critical point, these oscillations are superposed with substantial growth of the correlation lengths. The correlation length oscillates at double the frequency of the order parameter in the disordered phase, and matches the frequency of the order parameter oscillations in the ordered phase (see Methods). All data shown are for $\Omega/2\pi=3.1$ MHz.
}
\label{fig5}
\end{figure*}

\section*{Order parameter and ``Higgs'' oscillations}
\label{sec:amplitude}
In addition to the curvature-driven coarsening dynamics, our experiments clearly reveal persistent long-lived oscillations of the correlation length and the order parameter across a range of experimental parameters, as shown in Fig.~\ref{fig:correlations}c and Methods. We explore the origin of these oscillations in Fig.~\ref{fig5}. First, we apply local detunings to one of the two sublattices, which biases the order parameter. We then repeat the protocol described in Fig.~\ref{fig3:square defect}a, ramping to various values of $\Delta$ on both sides of the QPT. Directly after the ramp, we quench the pinning field off and follow the dynamics. We observe large-amplitude, long-lived oscillations of the order parameter, well modeled as a damped harmonic oscillator, $m_s(t) \approx \phi_0+A \cos{(\omega t+\theta_0)}e^{-\gamma t}$, with amplitude $A$, frequency $\omega$, damping $\gamma$, and offset $\phi_0$ that strongly depend on $\Delta$. We find that upon approaching the phase transition from both sides, $\omega$ decreases while $\gamma$ and $A$ increase; $\phi_0$ changes from zero in the disordered phase to a nonzero value in the ordered phase. 

To understand  the origin of these observations, we perform numerical simulations using matrix product state (MPS) methods. 
Through density-matrix renormalization group (DMRG) calculations on up to $10\times 10$ sites with periodic boundary conditions, we find that the pinned initial state corresponds to a low-energy state of the post-quench Hamiltonian even on the disordered side of the transition (Fig.~\ref{fig5}b). 
This is in contrast to the high-energy $\mathbb{Z}_2$ states typically associated with oscillations due to quantum many-body scars \cite{bernien2017probing,bluvstein2021controlling}.
We then attempt to simulate the dynamics of the pinned initial state using the time-dependent variational principle (TDVP) \cite{haegeman2016tdvp} at a relatively small bond dimension $\chi=256$ and find good qualitative agreement with the experimentally extracted frequencies (Fig.~\ref{fig5}d). 
Moreover, the oscillation frequencies closely match the numerically determined ground-state gap, and are relatively robust to variations in the system size away from the critical point (see Methods and ED Fig.~\ref{fig:higgs_supp}). 

In the ordered phase, these observed oscillations can be understood as an amplitude (``Higgs'') mode \cite{pekker2015amplitude}, which is a collective excitation of the magnitude but not the sign of the order parameter. Qualitatively, the ``Higgs'' mode can be viewed through the lens of Landau theory. In this framework, order-parameter dynamics are determined by an effective potential describing a quartic anharmonic oscillator: $V(\phi)=\frac{q}{2}\phi^2 + \frac{\lambda}{4}\phi^4 + \mathcal{O}(\phi^6)$, as shown in Fig.~\ref{fig5}c (we continue to identify the oscillation on the disordered side as a ``Higgs mode" for convenience). The ordered phase $(q\sim \Delta_c-\Delta <0)$ is differentiated from the disordered phase $(q>0)$ by a finite value $\phi_0\neq 0$ of the potential minima, which determines the offset of the oscillation. In the disordered phase, these are oscillations of the sign of the order parameter, hence they do not have the symmetry of a ``Higgs'' mode. Beyond the changing offset $\phi_0$, this simplified picture reproduces the increase of the oscillation amplitude $A$ and the decrease of the frequency $\omega$ when approaching the phase transition. To further investigate this amplitude mode, we prepare lower-energy biased states by softening the pinning field (smaller $|\delta_i|$). We find a corresponding decrease in the oscillation amplitude $A$ and the frequency $\omega$ with a sharper dependence near the QPT (Fig.~\ref{fig5}d, ED Fig.~\ref{fig:higgs_amplitude}), in agreement with Landau theory. The oscillation frequency of these lower-energy states shifts close to the many-body gap. In order to explore the ``Higgs'' mode deep in the ordered phase, where our state preparation scheme generates low-amplitude oscillations, we use an alternative experimental protocol, as detailed in the Methods and shown in ED Fig.~\ref{fig:ordered_quenches}.

These order parameter oscillations present a unique probe of the quantum critical point. In particular, the ratio of the oscillation frequencies on the two sides of the QPT is universal in equilibrium, and predicted to be $\omega(-|q|) / \omega(|q|)=\sqrt{2}$~\cite{sachdev2009exotic} by Landau mean-field theory. However, our experimental results, in which $\omega(-|q|) / \omega(|q|)>\sqrt{2}$ (see Extended Data Fig.~\ref{fig:higgs_ratio}), indicate a significant deviation from this simplistic prediction, broadly consistent with more advanced calculations predicting $\omega(-|q|) / \omega(|q|)\approx 1.9$ (see discussion in Methods). The discrepancy with mean-field results emphasizes the central role of quantum fluctuations, and in particular, finite-momentum order parameter fluctuations, in the vicinity of the QPT. The dynamics of these fluctuations are also expressed in the progressively larger oscillations of the correlation length (Fig.~\ref{fig5}e) and a sharp increase in the damping term $\gamma$ (Fig.~\ref{fig5}d) observed upon approaching the critical point.

\section*{}

\section*{Discussion and outlook}
\label{sec:outlook}
Our observations shed new light on paradigmatic collective processes in closed nonequilibrium quantum many-body systems,
highlighting the important role of coarsening dynamics, and revealing their curvature-driven character in systems with a nonconserved order parameter \cite{Bray1994,Samajdar2024}. Crucially, we measure an acceleration of the ordering processes when approaching the phase transition, a signature of the intrinsically quantum nature of the dynamics \cite{Samajdar2024}. While we observe a scaling collapse of the structure factor suggestive of self-similarity~\cite{Bray1994}, the dynamically varying amplitude $b(t)$ deviates from the expected universal behavior. As discussed in the Methods, this deviation could originate from finite-size effects (as well as, potentially,  residual disorder or decoherence), and its detailed understanding constitutes  an interesting theoretical problem \cite{barenblatt1996scaling}. Similar mechanisms may account for the slowdown of coarsening at late times observed in Figs.~\ref{fig:correlations}c,d.  Further evidence for the role of finite-size effects is provided by measurements involving local control: we find that the dynamics of domains seeded away from the system's boundaries (Fig.~\ref{fig3:square defect}) are in much closer agreement with  universal theoretical predictions. 

Additionally, we observe the concurrent excitation of the ``Higgs'' mode upon crossing the phase transition. Investigation of this mode yields detailed information on important observables---such as the damping rate of the order parameter/``Higgs'' mode in the vicinity of the phase transition---which are difficult to access classically \cite{sachdev1997theory}. For the numerically accessible system sizes and bond dimensions, our simulations cannot capture the damping rate $\gamma$, and generally break down near the critical point (see Methods). More generally, the possible interplay of coarsening with the ``Higgs'' mode presents an intriguing question that warrants further theoretical investigation. 

These studies can be extended along several directions. 
Several recent experiments demonstrated universal dynamics far from equilibrium, often interpreted through the framework of non-thermal fixed points \cite{schmied2019non}, including observations of simultaneous IR and UV scaling laws \cite{glidden2021bidirectional,gazo2023universal} and classification of universality classes \cite{huh2024universality}. Similar phenomena can be explored in non-mean field systems near QPT using  programmable simulators. Conversely, manifestations of quantum criticality and ``Higgs'' modes in non-equilibrium Bose gases may also be intriguing to explore.
In contrast to traditional condensed-matter systems, programmable quantum simulators can directly access correlation functions of \textit{any} order \cite{semeghini2021probing} as well as other important observables, such as the entanglement entropy \cite{brydgesProbingRenyiEntanglement2019,teng2024learning}, using, e.g., hybrid digital-analog approaches \cite{bluvstein2022quantum,andersen2024thermalization}. These could provide further insights into complex  dynamics, particularly near the quantum critical point, where numerical calculations are prohibitively challenging. Besides the symmetry-broken ordered states probed in this work, it would also be interesting to extend our study of coarsening dynamics to the formation of topologically ordered states of matter \cite{semeghini2021probing,Samajdar.2021,Verresen.2020,ott2024probing}, which cannot be characterized by local order parameters. Additionally, local programmability may be used to explore the tunneling of metastable states, known as false vacuum decay \cite{zenesini2024false, darbha2024false,darbha2024long}.

During the completion of this work, we became aware of related work demonstrating coarsening phenomena driven by quantum fluctuations on a superconducting quantum processor 
\cite{andersen2024thermalization}.

\clearpage
\newpage

\section*{Methods}

\noindent\textbf{Experimental platform}\\
A detailed description of our experimental platform is given in \cite{ebadi2021quantum, bluvstein2024logical}. All measurements are realized using a two-dimensional programmable quantum simulator based on Rydberg atom arrays. Single $^{87}$Rb atoms are stochastically loaded into optical tweezers shaped by a spatial light modulator (SLM), and then rearranged into defect-free patterns using a pair of crossed acousto-optic deflectors (AODs). Both sets of tweezers use 852\,nm-wavelength light. The atoms are then laser-cooled and optically pumped to the $\ket{5S_{\frac{1}{2}},F=2,m_F=-2}$ state, which we denote as $\ket{g}$ in the main text. A pair of counterpropagating lasers at 420 nm and 1013 nm wavelengths couple $\ket{g}$ to the highly excited Rydberg state $\ket{r} = \ket{70S_{\frac{1}{2}},J=\frac{1}{2},m_{J}=-\frac{1}{2}}$ via a two-photon transition through the $6P_{\frac{3}{2}}$ orbital, blue-detuned by approximately $2\pi\times 2.4$\,GHz with respect to the $5S_{\frac{1}{2}}\rightarrow6P_{\frac{3}{2}}$ transition.

We use $16\times 16$ lattices of atoms, and maintain an $R_b/a$ ratio of $1.12-1.15$, such that only nearest neighbors lie within the blockade radius. Due to the rapid fall-off of the van der Waals term, only nearest and next-nearest neighbor interactions meaningfully contribute to our observations. The data shown in the main text is taken with two-photon Rabi frequencies of either $\Omega/2\pi=3.8,~6.0,$ or $3.1$ MHz, with corresponding lattice spacings $a = 6.8,~6.45 $, or $ 7.15$\,$\mu m$ and $R_b/a = 1.15,~1.12,$ or $1.13$. The 1013\,nm and 420\,nm single-photon Rabi frequencies are approximately balanced $(\Omega_{1013}\approx\Omega_{420})$. The experiment time $T$, sweep parameters, and lattice spacings are all rescaled for different $\Omega$, such that $\Omega/\Delta$, $R_b/a$, and the total phase accumulated $\Omega T$ are constant when comparing different experimental configurations.

In order to observe coarsening, the sweep rate through the critical point must fall within a certain range. A sweep rate that is too slow  would create a low-energy state, and consequently, the coarsening dynamics may be too slow to measure. In contrast, a very fast sweep  or an instantaneous quench could inject too much energy and bring the system out of the ordered phase (which persists up to a finite energy density). Here, all measurements use linear sweeps with sweep rates of $\frac{\Delta/\Omega}{\Omega T}\approx\frac{1}{2\pi}\times 3.0$, which, we find, falls within the desired range.

For the global sweeps, we apply post-selection based on the success of the rearrangement protocol, selecting shots with $\leq 4$ defects. This threshold retains an average of 93\% of shots over all end detuning values presented in Fig.~1 and Fig.~2a,b. Additionally, we post-select on measurement results in order to discard runs in which we suspect large-scale errors have occurred. Due to the high energy of the Rydberg blockade, a large number of blockade violations are extremely unlikely to be naturally generated by the coherent dynamics of Eq.~\eqref{eq:ryd_ham}. Furthermore, since our projective measurement cannot differentiate $\ket{r}$ occupation from loss induced by other mechanisms,  we attribute the presence of a large number of apparent blockade violations to manifestations of unwanted noise processes, such as blackbody-induced avalanche decays \cite{festa2022blackbody}. We therefore discard runs where the longest chain of consecutive atoms in a single row or column detected in state $\ket{r}$ has a length of more than four sites. Small scale ED numerical simulations support that the probability of reaching such states from purely unitary dynamics is negligibly small. Imposing this post-selection threshold retains $92\%$ of the data. Imposing both rearrangement and avalanche post-selection, we retain 86\% of the global-sweep shots across all end-detunings presented.
We note that the post-selection is most significant for  the longest times and largest detunings presented ($t = 1.16 \mu s, \Delta/\Omega = 4$ in ED Fig.~2d), as larger  number of shots are corrupted through avalanche decays and we only retain $63\%$ of the shots at the longest time-step.
\\

\noindent\textbf{Local control}\\
To enable individual single-site addressing of atoms with a local light shift, we use an SLM (Hamamatsu LCOS-SLM X15213-02) to generate optical tweezers in arbitrary spatial patterns with beam waist $1\,\mu$m, ensuring robustness to atomic position fluctuations (Extended Data Fig.~\ref{fig:localcontrol_supp}). The wavelength we choose to operate at, 784\,nm, achieves a measured differential AC Stark shift between the $5S_{1/2}$ and $5P_{3/2}$ states of 12.2(3)\,MHz with $\approx160\,{\mu}$W per spot, but a negligible scattering rate ($\approx 35$ Hz) (the scaling of the light shift with laser amplitude is shown in Extended Data Fig.~\ref{fig:localcontrol_supp}c). The light is linearly polarized to minimize vector light shifts on the ground state hyperfine manifold. We further measure the shift on the $\ket{g}\rightarrow\ket{r}$ transition and find that the light shift is well approximated by the differential $5S_{1/2}\rightarrow5P_{3/2}$ light shift as $\delta_0 = -2\pi\times12(2)$\,MHz at the same power per spot. 

The phase holograms for the SLM are generated using the phase-fixed weighted Gerchberg–Saxton (WGS) algorithm, taking into account the desired position and relative intensity of the local light-shift pattern \cite{dongyukimGS}. We first generate a local addressing pattern that closely matches the positioning of the atomic tweezer array; however, perfect matching of the two arrays is computationally expensive as it requires an extremely high sampling rate of the image plane of the local addressing pattern. In order to overcome this computational barrier, after creating an initial local addressing pattern, we align it to the atom positions by transforming the phase hologram. By stretching, rotating, and applying tilts and defocus, we can match the two patterns with feedback on the atom signal. The latter three can be easily  controlled using Zernike polynomials, while the stretching and rotation require more care to preserve the intensity homogeneity of the desired pattern. We find that na\"ive rescaling or rotating of the hologram results in unwanted distortion of the intensity pattern, attributed to software interpolation when working with a pixelated hologram. This is mitigated by applying the computational corrections in the image plane. We take the Fourier transform of the hologram, convolve the intensity profile with a 2D Gaussian to broaden each spot over several pixels (in order to minimize effects of interpolation), and then apply the rotation and stretching. Lastly, we apply an inverse Fourier transform back to the Fourier plane and use the resultant phase hologram for the SLM. 
Using this procedure, we first coarsely align the individual addressing pattern to the tweezers on a camera, and then precisely align the two using a spin-echo measurement of the light shift (Extended Data Fig.~\ref{fig:localcontrol_supp}b) to optimize the alignment parameters such that the intensity is maximized at the atom sites. Good alignment is also crucial to prevent atom loss from turning on a misaligned potential.
Finally, we correct the tweezer intensities as required  using the fitted light shifts to feedback on the target weights in the hologram generation. 

Examples of states prepared using such tweezer profiles, where the detuning is used to strongly pin atoms to the ground state, are shown in Extended Data Fig.~\ref{fig:localcontrol_supp}d. At the boundary between different AF orders and the edges of the array, the mean-field repulsive interaction strength decreases for sites with fewer Rydberg neighbors; we therefore weight the local detuning strength inverse-proportionally to the number of neighbors. Note that when arbitrary weighting is used, the total power remains constant (number of addressed sites $\times~2\pi\times12$ MHz), but the power is redistributed in the tweezers accordingly. Nevertheless, particularly at large $\Delta/\Omega$, neighboring Rydberg excitations start to be energetically favored (antiblockaded) along the domain boundaries. Excluding such edge effects, the preparation probability of preparing the single-atom ground state on the pinned sites is 93-95\% (Extended Data Fig.~\ref{fig:localcontrol_supp}e).

For other realizations of single-site addressing using light shifts in atom arrays, see e.g. \cite{chen2023continuous, de2024demonstration}.\\ 

\noindent\textbf{Theoretical background of coarsening dynamics}\\
In this section, we summarize the theoretical details of the different kinds of coarsening processes that govern the dynamics of the system as long-range order is formed. Although our focus will be on the Rydberg atom array, to begin, let us consider the generic situation of a  system driven through a continuous quantum phase transition (QPT) by tuning some parameter of the Hamiltonian, $g$,  linearly with time. Without loss of generality, we assume that the quantum critical point (QCP) is located at $g=0$ and the zero of time is set such that $g(t) = t/\tau$; hence, the system crosses the QCP at $t=0$. For the specific case of the neutral atom array considered in this work, the time-dependent parameter $g$ can be defined as $g (t) = (\Delta (t) - \Delta_c)/\Omega$.

As the system approaches the quantum critical point, its relaxation time diverges and it necessarily falls out of equilibrium. However, \textit{when} it does so depends on the velocity of the linear ramp, $\dot{g} (t) = 1/\tau$.
The quantum Kibble-Zurek mechanism posits that the time at which the system's evolution ceases to be adiabatic is $t$\,$=$\,$-t_{\textsc{kz}}$ with $t_{\textsc{kz}} \sim t_0(\tau/t_0)^{\nu z/(\nu z+1)}$, where $\nu$ is the correlation length exponent, $z$ is the dynamical critical exponent, and $t_0$ is some microscopic time scale. Thereafter, since the system cannot dynamically respond fast enough to the changing parameter of the Hamiltonian, it remains ``frozen'' through a so-called impulse regime until a later time $t=+t_{\textsc{kz}}$, when it unfreezes on the other side of the quantum phase transition. During this impulse regime, the KZM presumes that the system's correlation length remains the same as when it initially froze: $\xi_{\textsc{kz}} \sim l_0(\tau/t_0)^{\nu/(\nu z+1)}$, where $l_0$ is some microscopic length scale. As a consequence, in this picture, the correlation length in the ordered phase is also set by $\xi_{\textsc{kz}}$ with no subsequent dynamics.

However, the nonequilibrium correlation length of the system, $\xi (t)$, can and does grow both in the impulse regime as well as in the ordered phase since the long-range correlations take time to develop. In the experiments described in the main text, this occurs via a two-step process.
First, as the system passes through the quantum critical regime, it undergoes \textit{quantum critical coarsening}, which is governed by the dynamical critical exponent $z$ of the particular QCP; for the $(2+1)$D Ising transition, $z=1$. Then, as time progresses and the ramp continues, the system eventually enters the ordered phase. Here, once the growing nonequilibrium correlation length $\xi(t)$ exceeds the equilibrium correlation length of the quantum ground state (which, recall, scales as $\xi_q \sim \lvert g \rvert^{-\nu}$), the dynamics cross over to a regime of \textit{noncritical coarsening}, for which
\begin{equation}
\label{eq:expect}
 \frac{d\,\xi(t)}{dt} \sim \frac{\xi_q^{z_d}~\varepsilon}{(\xi(t))^{z_d-1}},
\end{equation}
where $\varepsilon$ is the many-body gap between the ground and first-excited states.  The dynamical exponent $z_d$ is dependent on the dimensionality and conservation laws of the system. For curvature-driven coarsening dynamics with a nonconserved scalar order parameter---as is indeed the case experimentally---$z_d = 2 > z$. A particular feature of noncritical coarsening worth emphasizing is the dependence of the dynamics on the distance to the QCP encoded in  Eq.~\eqref{eq:expect}. Specifically, the ground-state equilibrium correlation length scales as $\xi_q \sim \lvert g \rvert^{-\nu}$ and the gap $\varepsilon \sim \lvert g \rvert^{\nu z}$. Plugging in the exponents of the $(2+1)$D Ising QPT, $\nu = 0.629$ and $z=1$, along with $z_d = 2$, in Eq.~\eqref{eq:expect}, we find the growth law
\begin{equation}
 \frac{d\,\xi(t)}{dt} \sim \frac{(\Delta - \Delta_c)^{-0.629}}{\xi(t)},
\end{equation}
This relation can be observed in Fig.~\ref{fig3:square defect}, which studies the rate at which a locally introduced domain in the center of the array shrinks.
The area of such a domain decreases at a rate $d r^2/dt \sim - \xi d \xi/d t$, which scales as $(\Delta - \Delta_c)^{-0.629}$, in consistency with the behavior observed in Fig.~\ref{fig3:square defect}d.

For a ramp that continues indefinitely without stopping, the entire dynamical evolution of the correlation length can be described by a single universal scaling function encompassing the adiabatic, quantum critical coarsening, and noncritical coarsening regimes \cite{Samajdar2024,biroli2010kibble,chandran2012kibble}: 
\begin{equation}
\label{eq:ansatz}
    \xi(t) \approx \xi_{\textsc{kz}} f\left(\frac{t}{t_{\textsc{kz}}} \right) \equiv \xi_{\textsc{kz}} f\left(x \right),
\end{equation}
where $f(x)$ is some universal function, and $\xi_{\textsc{kz}}$ and $t_{\textsc{kz}}$ depend on the ramp rate $\tau$ as specified earlier. The scaling variable $x$ delineates the three regimes discussed above as
\begin{alignat}{2}
\nonumber
    &x\ll -1: &&\mbox{ adiabatic},\\
    &|x|\lesssim \mathcal{O}(1): &&\mbox{ quantum critical coarsening},\\
    \nonumber
    &x\gg1:&& \mbox{ noncritical coarsening}.
\end{alignat}
More generally, if the ramp is stopped at a time $t_s$, the dynamical scaling form is altered to
\begin{equation}
\label{eq:Fgen}
    \xi (t) \approx \xi^{}_{\textsc{kz}}\, \mathcal{F} \left( \frac{t}{t_{\textsc{kz}}}, \frac{t_s}{t_{\textsc{kz}}}\right) \equiv \xi^{}_{\textsc{kz}}\, \mathcal{F}(x,x_s)~;
\end{equation}
for $x\leq x_s$, one recovers the earlier scaling as $\mathcal{F}(x,x_s)=f(x)$.
If the ramp is stopped at $x_s=t_s/t_{\textsc{kz}}\gg 1$ in the noncritical coarsening regime, the behavior of the universal scaling function  for $x>x_s$ describes the physics during the hold time and is given by
\begin{equation}
\label{eq:SFy2}
\mathcal{F}(x, x_s) \approx
    x_s^{-\nu+(\nu z/z_d)}\left(\mathcal{C}x-\mathcal{C}_s x_s\right)^{1/z_d}~, 
\end{equation}
for some $\mathcal{O}(1)$ constants $\mathcal{C}$\,$>$\,$\mathcal{C}_s$.  Note that because $z$\,$<$\,$z_d$, the coarsening speeds up as we stop earlier in the ordered phase, closer to the QCP (which results in a lower $x_s$). This is indeed what we observe experimentally during the hold time following global sweeps across the phase transition, as shown in the inset of Fig.~\ref{fig:correlations}c. Intuitively, this is because a smaller $\Delta/\Omega$ corresponds to a greater relative influence of critical coarsening, which is faster than noncritical coarsening. 

In contrast, near the thermal phase boundary, the system can undergo an interval of \textit{classical critical coarsening}, which is described by a growth law $\xi (t) \sim t^{1/\bar{z}}$ with a distinct dynamical exponent. For the 2D classical Ising phase transition, $\bar{z} \approx 2.16 > z_d$ \cite{humayun1991non}, so the growth of correlations via classical critical coarsening is slower than for noncritical coarsening. Correspondingly, the dynamics should decelerate as one approaches  the classical critical point, in sharp contrast to the speedup outlined above in the vicinity of a QCP.\\

\noindent\textbf{Structure factor and correlation length}\\
In order to extract the structure factor and the correlation length \cite{sandvik2010computational}, we first calculate the two-point connected correlation function $G(\textbf{r}_1,\textbf{r}_2) =  \braket{\tilde{Z}_{\textbf{r}_1}\tilde{Z}_{\textbf{r}_2}}-\braket{\tilde{Z}_{\textbf{r}_1}}\braket{\tilde{Z}_{\textbf{r}_2}}$ and then average over all pairs of points with identical displacements $\textbf{r}$:
\begin{equation}
    G(\textbf{r}) = \frac{\sum_{\textbf{r}_1,\textbf{r}_2} G(\textbf{r}_1,\textbf{r}_2) \,\delta_{\textbf{r}_1-\textbf{r}_2,\textbf{r}}}{\sum_{\textbf{r}_1,\textbf{r}_2}\delta_{\textbf{r}_1-\textbf{r}_2,\textbf{r}}}.
\end{equation}
We first derive the standard structure factor by computing the Fourier transform of $G(\textbf{r})$,
\begin{equation}
    S(\textbf{k}) \equiv \mathcal{F}_k[G(\textbf{r})] = \sum_{\textbf{r}} e^{-i \textbf{k}\cdot \textbf{r} }G(\textbf{r}),
\end{equation}
and then calculate the radially averaged structure factor,
\begin{equation}
S(k)=\frac{\sum_{\textbf{k}}S(\textbf{k})\,\delta_{|\textbf{k}|,k}}{\sum_{\textbf{k}}\delta_{|\textbf{k}|,k}}.
\end{equation}
To extract a correlation length, we fit $S(k)$ to:
\begin{equation}
    S(k)\approx \frac{S_0}{(1+\xi^2 k^2)^{\frac{3}{2}}},
    \label{eq:sf_func}
\end{equation}
and we factorize $S_0$ as $S_0= b \xi^2/\pi$.
This form of Eq.~\eqref{eq:sf_func} is equivalent to assuming that the position-space correlations follow an exponential decay, $G(r)\approx A\exp({-\frac{r}{\xi}})$ \cite{ebadi2021quantum},  up to finite-size corrections. 
While equilibrium considerations for an infinite-size system suggest that for small $k$, $S(k)$ should obey the Ornstein-Zernike form $S(k)\approx \frac{S_0}{1+\xi^2 k^2}$ \cite{kennedy1991ornstein}, we empirically find that Eq.~\eqref{eq:sf_func} better captures our observed nonequilibrium distributions.

For universal coarsening dynamics, we theoretically expect $b$ to be constant. While our data indeed exhibit scaling collapse as in Eq.~\eqref{eq:sf_func}, we find that
$b$ varies during the dynamics, indicating the presence of an additional length scale(s). We observe that $b$ is correlated with $\xi$ and depends on $\Delta/\Omega$, as shown in ED Fig.~\ref{fig:collapse}c). Additional length scales that may affect the dynamics include the finite system size, the finite width of the domain walls (which depends on $\Delta/\Omega$), spatial inhomogeneity in $\Delta$ and/or $\Omega$, and length scales introduced by decoherence effects such as  decay due to the finite lifetime of the Rydberg state. 

Specifically, the expected universal scaling regime is expected to hold for distances $r$ and correlation lengths $\xi$ such that $l\ll r,\xi\ll L $, where $l$ is the width of a domain wall and $L$ is the system size \cite{Bray1994}, indicating that finite size effects are likely playing an important role in the present experiments. Hence, observing the theoretically expected universal coarsening behaviour in global quenches would likely require access to larger system sizes and correspondingly longer experiment times.
While recent experimental advances in neutral atom array platforms suggest that lattices more than an order of magnitude larger than the one presented in this paper are within reach \cite{manetsch2024tweezer}, elsewhere we describe the use of local control to deterministically nucleate and study domain dynamics away from the system's boundaries, allowing us to study universal properties of coarsening under present experimental conditions. \\

\noindent\textbf{Analysis of domains in global sweeps }\\
Using  single-site-resolved detection, we can map out the domains in each snapshot. First, we calculate the local staggered magnetization. Each domain is then identified as a region of the array where the same ordering $\textrm{AF}_1$ or $\textrm{AF}_2$ is connected by nearest neighbors. We do not consider single spins of opposite order as a separate domain. For Fig.~\ref{fig:global coarsening}a,b, we therefore first identify and correct individual spin flips. These are identified as single atoms which are of the opposite order compared to all of their nearest and next-nearest neighbors. Only after we have identified single spin flips and corrected them to match their surrounding bulk order do we identify the domain boundaries. A domain's area is defined as the total number of atoms comprising the domain. For the probability distribution of domain occupations presented in Fig.~\ref{fig:global coarsening}a, the frequency of each domain area is weighted by the area of that domain. We normalize the distribution by the sum of all area-weighted frequencies at each time step.\\

\noindent\textbf{Classical energy analysis}\\
To calculate the classical energy per single shot of the experiment, we first perform the single-spin-flip correction as described above. We then identify regions of the array which do not belong to either AF ordering by calculating a coarse-grained local staggered magnetization with a similar approach to previous works \cite{ebadi2021quantum}. In this work, specifically, we calculate the convolution $C_{x,y}$ of the Rydberg occupation $n_{x,y}$ with the kernel $ W = \big(\begin{smallmatrix}
  0 & 1 & 0\\
  1 & 0 & 1\\
  0 & 1 & 0
\end{smallmatrix}\big)$ for each snapshot. The output values of $C_{x,y}$ range from 0 to 4, where the extremal values correspond to atoms surrounded by nearest and next-nearest neighbors that all belong to the same AF orderings, as shown in Extended Data Fig.~\ref{fig:local_analysis}a. We consider an atom to be at a boundary if $n(x,y)= 1 (0)$ and $C_{x,y}\neq0(4)$ (see Extended Data Fig.~\ref{fig:local_analysis}b). In the raw array (not single-spin-flip corrected), we then compute the classical energy using Eq.~\eqref{eq:Classical_Hamiltonian} for each snapshot. 
The value of the interaction energy is calculated from the lattice spacing ($a$) and $V_0$ as $V_{nn} = V_0/a^6$. For the dataset presented  in Fig.~\ref{fig:global coarsening}c, $V_{ij} = V_{nn}/2\pi = 11.69$\,MHz for $\Omega/2\pi = 6$\,MHz. We  also account for next-nearest neighbor contributions. For each next-nearest neighbor Rydberg atom per snapshot, an additional $V_{nnn}/2\pi = 1.46$\,MHz is considered in the classical energy. The effects of longer range interactions are negligible. By using the spin-flip-corrected lattice for domain identification and the uncorrected one for the subsequent energy calculation, single spin flips that occur contribute to the classical energy in identified domain walls and bulk orderings but not as separate domains. We exclude the layer of atoms closest to the edge of the array for all contributions to the classical energy. Note that for the classical energy calculation in Fig.~\ref{fig:global coarsening}c, we post-select such that the maximum number of directly adjacent Rydberg atoms can be no more than three (compared to four used throughout this work). Due to the sensitivity of the boundary identification procedure used here to correlated decays, this post-selection is slightly stricter. In the data presented in Fig.~2c, as the data is at high end-detuning, we retain 69\% of data due to the avalanche post-selection and 84\% due to rearrangement post-selection. Overall, we retain 58\% of data.\\

\noindent\textbf{Analysis of locally prepared domains}\\
We estimate the local radius of curvature $R$ given in the equation $\partial_tR^2 = -v_a$ as the radius, $r$, of the central domain in Fig.~\ref{fig3:square defect}. The radius $r$ is defined as the Manhattan distance, $d_m$, at which the radially averaged local staggered magnetization, $m_s$, crosses zero (see Extended Data \ref{fig:local_analysis}c). We consider Manhattan distances instead of Euclidian distances from the center of the injected domains as the former is more representative of the nearest-neighbor interactions dominating the dynamics for very short times (on the order of one Rabi cycle). For long times, both measurements of distances in our lattice reveal a collapse to the linear form shown in Fig.~\ref{fig3:square defect}c. For this analysis, we only consider the unpinned sublattice. The local state-preparation protocol prepares states where atoms that are locally detuned are prepared in $\ket{g}$ with high probability for all values of $\Delta/\Omega$ (Extended Data Fig.~\ref{fig:localcontrol_supp}d). We observe that for $\Delta/\Omega$ close to the QPT, there is a larger discrepancy of the local order parameter $m_s$ between the pinned and unpinned sublattices. Therefore, when considering both sublattices, the radius is less clearly defined by a single point at which the order parameter crosses zero. We fit a linear relationship to extract $dr^2/dt$ at each $\Delta/\Omega$. 

A similar procedure is followed for the analysis of the coarsening dynamics in the zigzag domain wall in Fig.~\ref{fig4:zigzag}. Here too, the mean value of the local order parameter $m_{x,y}$ is calculated at each lattice site per time step. For Fig.~\ref{fig4:zigzag}c, the domain wall's horizontal position is calculated as the point at which the linearly interpolated line between points crosses $m_s = 0$. In Extended Data Fig.~\ref{fig:local_analysis}d, we show the variation of the domain-wall position with hold time for two additional values of $\Delta/\Omega$. These data points reinforce the strong $\Delta/\Omega$-dependence of the domain-wall velocity already seen in Fig.~\ref{fig3:square defect}c, d. Note that for this analysis, we include atoms which were initially locally detuned as we are considering each row separately. We therefore see larger uncertainty in the domain walls' positions for low $\Delta/\Omega$ (see Extended Data Fig.~\ref{fig:local_analysis}d). 

Errors in Figs.~\ref{fig3:square defect}c, \ref{fig4:zigzag}c, and Extended Data Fig.\ref{fig:local_analysis}d are calculated using bootstrapping. From the full set of experimental single snapshots of size $N\sim600$ shots, we sample $N$ times with replacement and calculate the value of interest ($r^2$ or the horizontal domain-wall position $x$) on each sample. The plotted errorbar is the standard deviation of the value of interest, calculated from 1000 repetitions of the above procedure. \\

\noindent\textbf{Numerical simulations of local domains}\\
We simulate the dynamics of locally prepared domains using the time-dependent variational principle (TDVP) \cite{haegeman2011tdvp,haegeman2016tdvp}. We use a two-site variant of this algorithm, which allows the bond dimension to grow with the evolution time at the expense of forgoing strict energy conservation due to the truncation step involved. In our calculations, we find that the energy is conserved to within $0.004 \%$ of that of the initial state up to the longest times simulated.

The initial states for the numerics are chosen to be a mean-field approximation of the experimentally prepared state. Specifically, we pin certain lattice sites to $\rvert g \rangle$, as specified by the target configuration, while the remaining ones are set to the vector on the Bloch sphere that minimizes the system’s mean-field energy for a given $\Delta/\Omega$ (instead of the fully polarized $\rvert r \rangle$ state). The many-body evolution is simulated using a maximum bond dimension of $\chi = 1200$ with a time step $\Delta t$ of $0.2 \Omega^{-1}$ (the dynamics are also consistent with those for a smaller $\Delta t = 0.1 \Omega^{-1}$).
The results thus obtained are showcased in Extended Data Fig.~\ref{fig:local_numerics} and are found to be in good agreement with the experiments.\\

\noindent\textbf{Amplitude/``Higgs'' mode}\\
\\
\noindent{\it Background and theory} \\
\\
To describe the observed amplitude mode, we consider the low-energy effective action that describes the transition between the disordered and antiferromagnetic phases. Its Lagrangian is the $\phi^4$-theory
\begin{align} \label{eq:lagrangian}
    {\cal L}[\phi] = \frac{1}{2} \Big[ (\partial_t \phi)^2 + (\nabla \phi)^2 - q \phi^2 \Big] - \frac{\lambda}{4}\phi^4, 
\end{align}
where $\phi$ corresponds to the coarse-grained order parameter of the antiferromagnetic phase.
While the phase transition 
is described via the Wilson-Fisher fixed point, here we perform a simple mean-field treatment of Eq.~\eqref{eq:lagrangian} in order to capture the physics away from the immediate vicinity of the transition.
In particular, the classical mean-field equation of motion for the order parameter's expectation value is given by
\begin{equation} \label{eq:op_eom}
    \partial_t^2 \phi = - (q+\lambda \phi^2) \phi,
\end{equation}
which corresponds to a classical anharmonic oscillator. The stationary value of the order parameter is given by $\phi_0=0$ in the disordered phase ($q>0$) and $\phi_0 = \pm \sqrt{-q/\lambda}$ on the ordered side of the transition ($q<0$). 
Expanding Eq.~\eqref{eq:op_eom} for small amplitudes, $\phi = \phi_0 + \delta \phi$ ,around the potential minima leads to harmonic oscillations of the order parameter with frequencies $\omega(q>0) = \sqrt{q}$ and $\omega(q<0) = \sqrt{2|q|}$.\\

\noindent{\it Numerical simulations} \\
\\
We first investigate the low-energy spectrum of the 2D Rydberg Hamiltonian at different values of $\Delta/\Omega$ using the excited-state density matrix renormalization group (DMRG) method, which iteratively finds the lowest-energy eigenstate that is orthogonal to previous lower-energy eigenstates. Our simulations take into account van der Waals interactions up to third-nearest neighbors on the square lattice.
We identify the gapped paramagnetic and spontaneous-symmetry-breaking (SSB) phases, and obtain the ground-state energy as well as the energy gaps $\Delta E_1$, $\Delta E_2$ of the first two excited states above the ground state (in the SSB phase, the first excited state that we identify is the symmetry-related ground state). 
In Extended Data Fig.~\ref{fig:higgs_supp}a, we perform bond-dimension scaling for  $\Delta E_1$, $\Delta E_2$ on a $10 \times 10$ lattice with open (OBC) and periodic (PBC) boundary conditions up to bond dimensions $\chi=512$ and $\chi=256$, respectively. We note that while the energy gap in the paramagnetic (disordered) phase is robust to boundary conditions, in the SSB (ordered) phase, we identify a distinct boundary energy gap (panel i) smaller than the bulk energy gap (panel ii). In the dynamics, the coupling of the order parameter to the boundary mode vanishes with increasing system size, and we thus consider the bulk gap extracted from periodic boundary conditions as the relevant frequency of the amplitude (Higgs) mode.

Furthermore, we perform DMRG calculations to obtain the initial state of the quench dynamics shown in Fig.~\ref{fig5}, which is the ground state of the Hamiltonian with additional local detunings $|\delta_l|/\Omega=0.7$ that pin one sublattice of the checkerboard to the ground state. We evaluate the energy expectation value of this initial state with respect to the unpinned 2D Rydberg Hamiltonian, and compare this with the energy of a $\mathbb{Z}_2$ product state, confirming that the pinned state is indeed a low-energy state (see Fig.~\ref{fig5}b).

Dynamics from the low-energy pinned initial state lead to amplitude oscillations at frequencies matching the ground-state gap, which we simulate using TDVP \cite{haegeman2011tdvp,haegeman2016tdvp} on a $10 \times 10$ lattice at a bond dimension $\chi=256$. 
We use time steps of $0.25 \Omega^{-1}$ and have verified that the resulting dynamics agrees with smaller time steps of $0.1\Omega^{-1}$. 
As seen in Extended Data Fig.~\ref{fig:higgs_supp}b, for small systems with OBC, the dynamics in the ordered phase (here, $\Delta/\Omega=2$) are dominated by a slow mode with frequency matching the boundary gap shown in Extended Data Fig.~\ref{fig:higgs_supp}a.i. 
On top of this slow mode, we see the presence of a mode with higher frequency matching the bulk gap shown in Extended Data Fig.~\ref{fig:higgs_supp}a.ii. Therefore, in the following, we consider dynamics with PBC, which allow us to isolate the bulk mode. 

In Extended Data Fig.~\ref{fig:higgs_supp}c, we observe clear oscillations of the order parameter deep in either phase, which we fit to a damped harmonic oscillator $\phi(t) \approx \phi_0+A \cos{(\omega t+\theta_0)}e^{-\gamma t}$ with frequency $\omega$, damping $\gamma$, offset $\phi_0$, and amplitude $A$. We show the dependence of these parameters on the ratio $\Delta/\Omega$ in Extended Data Fig.~\ref{fig:higgs_supp}d. Away from the phase transition, the oscillation frequencies overlap with the previously obtained ground-state energy gaps and are robust to system size. 
We further see that the damping and amplitude become larger towards the transition, where the offset acquires a nonzero value.
Close to the transition, the TDVP dynamics fails to converge in bond dimension and fit to the damped harmonic oscillator's functional form, as apparent in Extended Data Fig.~\ref{fig:higgs_supp}c.ii. Moreover, even away from the critical point, the limited bond dimension does not capture a finite damping rate of the oscillations.\\

\noindent{\it Dependence on local detuning strength} \\
\\
We experimentally investigate the dependence of the amplitude mode on the strength of the applied local detunings at two points, $\Delta/\Omega = 0$ and $\Delta/\Omega=1.1$. This data is obtained by performing the state-preparation sequence illustrated in Fig.~\ref{fig3:square defect}a for varying strengths of the applied local Stark shift $\delta$. For all other uses of the local detunings in this work, the pinning is applied at a constant magnitude, $\delta_0 = -2\pi\times12(2)$MHz, and the corresponding state preparation for various $\Delta/\Omega$ is documented in Extended Data Fig.~\ref{fig:localcontrol_supp}c, d. Here, this strength is varied to much lower values than for the saturated pinning of the ground state ($0.04\delta_0-0.1\delta_0$) in the rest of the work (as indicated in Extended Data Fig.~\ref{fig:ordered_quenches}a,b) before significant differences in the resultant oscillations are observed.

We find that at both values of $\Delta/\Omega$ considered, the amplitude of the oscillations is progressively reduced as $|\delta|$ is decreased. The frequency of the oscillations also decreases with $|\delta|$. The change in the oscillation frequency is more pronounced for oscillations near the critical point than in the disordered phase. The $\delta$ dependence of the oscillation frequency, as well as the increased sensitivity near the phase transition, are qualitatively consistent with the behavior of an anharmonic oscillator, as predicted by the mean-field Eq. \eqref{eq:op_eom}.\\

\noindent{\it Amplitude mode in global sweeps} \\
\\
The ``Higgs''-mode dynamics are also apparent in parallel to coarsening, as manifested in the oscillations of the magnetization $n_i$ (Extended Data Fig.~\ref{fig:higgs_global}a). Additionally, they can be clearly discerned by observing the dynamics of the total magnetization of the two-point correlation function in position space $C(r)$, as shown in Extended Data Fig.~\ref{fig:higgs_global}b--d. The frequency of the oscillations of the correlation length closely support those extracted from the quench protocol described below (Extended Data Fig.~\ref{fig:ordered_quenches}) and the calculated ground-state energy gap (the global sweep data is plotted in purple in Fig.~\ref{fig5}). As mentioned earlier, the interplay of the amplitude mode with coarsening dynamics is generally unexplored theoretically. Therefore, we note further exploration of the two processes occurring in parallel, as a possible future extension of this work. \\

\noindent{\it Quenches in the ordered phase}\\
\\
Although ``Higgs'' oscillations in the ordered phase can be extracted through the state-preparation sequence through deterministic preperation with local detunings, as described in Fig.~\ref{fig3:square defect}a, the amplitude of the oscillations is substantially reduced when compared to that inside the disordered phase (see $\Delta/\Omega = 1.5$ in Fig.~\ref{fig5}). We therefore perform an alternate state-preparation sequence to extract the amplitude-mode frequencies deep in the ordered phase, as shown in Extended Data Fig.~\ref{fig:ordered_quenches}a. First, local detunings are applied in a checkerboard pattern  as the global detuning $\Delta$ is swept from negative values to $\Delta/\Omega = 3.3$, a point far inside the ordered phase. We hold the global detuning constant while quenching off the site-dependent $\delta$. At this point, we quench $\Delta$ to its final detuning value in the ordered phase, yet closer to the phase transition. By way of this protocol, we observe, as with all other sweep protocols presented thus far, long-lived oscillations of the order parameter and correlation lengths as shown in Extended Data Fig.~\ref{fig:ordered_quenches} b,c. The extracted oscillation frequencies $\omega$ are in close agreement with the ground-state energy gap in the ordered phase. Note that in Fig.~\ref{fig5}d, the points in red at $\Delta/\Omega  = 2.0,~2.5$ are measured via this quench protocol. \\

\noindent{\it Frequency doubling}\\
\\
From the above-mentioned quenches to the ordered phase as well as in data from the local protocol, we extract the oscillations in both the order parameter and the correlation lengths. We find, as shown in Extended Data Fig.~\ref{fig:ordered_quenches}d, that in the ordered phase, these two frequencies are approximately equal, while in the disordered phase they vary by $\omega_{\xi}/\omega_{m_s} \approx 2$. The changing relationship between the two observables can be understood by the following symmetry argument. 

We begin by taking into account the dynamics of order-parameter fluctuations within a Gaussian approximation. Neglecting corrections to the effective mass due to fluctuations, the relevant equations of motion are~\cite{chandran2013equilibration,dolgirev2022periodic}:
\begin{alignat}{1}
    \partial_t {\cal D}^{\phi\phi}_{\bm k,t} &= 2{\cal D}^{\phi\pi}_{\bm k,t},\label{eqn:dtD1}\\
    \partial_t {\cal D}^{\phi\pi}_{\bm k,t} &= {\cal D}^{\pi\pi}_{\bm k,t}  
    - (k^2 + q +  3\lambda \phi^2){\cal D}^{\phi\phi}_{\bm k,t},\label{eqn:dtD2}\\
    \partial_t {\cal D}^{\pi\pi}_{\bm k,t} &= - 2(k^2 + q +  3\lambda \phi^2){\cal D}^{\phi\pi}_{\bm k,t}, \label{eqn:dtD3}
\end{alignat}
where $\pi_{\bm k}(t) \equiv \partial_t \phi_{\bm k}(t)$ and ${\cal D}^{\phi\phi}_{\bm k,t} \equiv \langle \phi({-\bm k,t})\phi({\bm k,t})\rangle_c$. From the correlation function ${\cal D}^{\phi\phi}_{\bm k,t}$, which corresponds to the structure factor discussed in the main text, one can extract the evolution of the correlation length.

Expanding $\phi = \phi_0 + \delta \phi$ and ${\cal D}^{\phi\phi}_{\bm k,t} = {\cal D}^{\phi\phi}_{\bm k} + \delta{\cal D}^{\phi\phi}_{\bm k,t}$, in the disordered phase, an eigenmode analysis of Eqs.~\eqref{eqn:dtD1}--\eqref{eqn:dtD3} yields a frequency spectrum $2\sqrt{k^2 + q}$. The smallest frequency, which is expected to set the correlation-length oscillations~\cite{dolgirev2022periodic}, is thus $2\sqrt{q}$, i.e., twice that of the order parameter.
In contrast, in the ordered phase, Eq.~\eqref{eqn:dtD2} contains a term $6\lambda \phi_0 {\cal D}_{\bm k}^{\phi\phi} \delta\phi(t)$, and the oscillation  of the order parameter thus acts as a linear drive on the dynamics of two-point correlation functions. 
As such, the correlation length will oscillate at the corresponding frequency $\sqrt{2|q|}$ of the order parameter.\\

\noindent{\it Frequency ratio of oscillations}\\
\\
In Extended Data Fig.~\ref{fig:higgs_ratio} we present the full dataset of amplitude-mode oscillations (also shown also in Fig.~\ref{fig5}) as a function of the distance from the phase transition.
As described above, Landau mean-field theory  predicts the relationship between oscillation frequencies to be $\omega(-|q|)/\omega(|q|) = \sqrt{2}$ \cite{ruegg2008quantum, sachdev2009exotic}. 
However, beyond mean-field theory, the phase transition is described by the Wilson-Fisher fixed point,
and the universal frequency ratio is shifted accordingly.
Theoretical estimates based on both analytical and numerical methods yield a frequency ratio around $\omega(-|q|)/\omega(|q|) \approx 1.9$~\cite{sachdev1997theory, brezin1974,caselle1997,campostrini2002}.
We find that both the experimental data and MPS simulations deviate from the mean-field prediction and are suggestive of a similarly higher ratio. However, very close to the critical point, $|(\Delta-\Delta_c)/\Omega|\lesssim 0.3$, we observe deviations from this theoretical ratio.
Possible explanations (besides the limited bond dimension for the MPS data) include finite-size effects (resulting e.g.  in non-vanishing gap at QPT point), possible errors in QPT location, and the possibility that  sufficiently close to the transition, the overdamped oscillations may no longer track the ground-state excitation gap.
In future work, a detailed exploration of the region near the critical point via the amplitude mode could allow for higher precision tests of this universal ratio. \\

\noindent\textbf{Data Availability}\\
The data that supports the findings of this study are available from the corresponding author on reasonable request.\\

\noindent\textbf{Acknowledgements}\\ 
We thank Manuel Endres, Tout Wang, Dries Sels, Hannes Pichler, Maksym Serbyn and Ahmed Omran for helpful discussions.
MPS simulations were performed using the TeNPy library~\cite{hauschild2018tenpy}.
We acknowledge financial support from the US Department of Energy (DOE Quantum Systems Accelerator Center, grant number DE-AC02-05CH11231, and DE-SC0021013),
the DARPA ONISQ program (grant number W911NF2010021), the DARPA IMPAQT program (grant number HR011-23-3-0030), the Center for Ultracold Atoms (an NSF Physics Frontiers Center), the National Science Foundation, 
and  QuEra Computing.
T.M. and J.F. acknowledge support from the Harvard Quantum Initiative Postdoctoral Fellowship in Science and Engineering. S.J.E. acknowledges support from the National Defense Science and Engineering Graduate (NDSEG) fellowship. D.B. acknowledges support from the NSF Graduate Research Fellowship Program (grant DGE1745303) and the Fannie and John Hertz Foundation.  N.U.K. acknowledges support from The AWS Generation Q Fund at the Harvard Quantum Initiative. N.M. acknowledges support by the Department of Energy Computational Science Graduate Fellowship under award number DE-SC0021110.  R.S. is supported by the Princeton Quantum Initiative Fellowship. D.A.H. was supported in part by NSF QLCI grant OMA-2120757. S.S. acknowledges support by U.S. National Science Foundation grant DMR-2245246.
\\

\noindent\textbf{Author contributions} \\
T.M., S.H.L., S.E., A.A.G., S.J.E., D.B., H.Z. and M.K. contributed to building the experimental setup, performed the measurements and analysed the data.
R.S., N.U.K., J.F., P.E.D., N.M., S.S and D.A.H. performed  numerical simulations and contributed to the theoretical prediction and interpretation of the results.
All work was supervised by S.S, D.A.H., M.G., V.V. and M.D.L. All authors discussed the results and contributed to the manuscript. 
\\

\noindent\textbf{Competing interests:} M.G., V.V., M.D.L. are co-founders and shareholders and H.Z. is an employee of QuEra Computing.\\ 

\noindent\textbf{Correspondence and requests for materials} should be addressed to M.D.L.\\

\setcounter{figure}{0}
\newcounter{EDfig}
\renewcommand{\figurename}{Extended Data Fig.}

\begin{figure*}
\centering
\includegraphics[width=2\columnwidth]{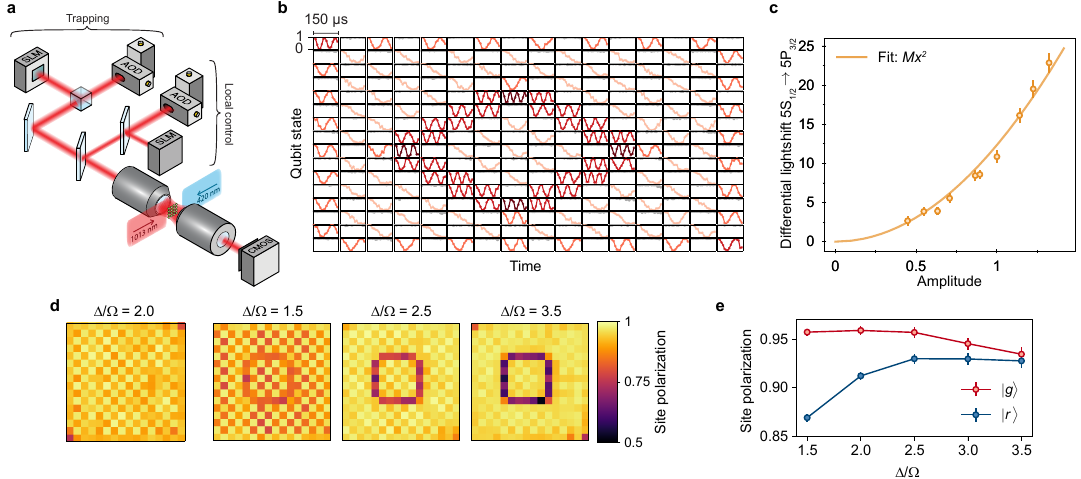}
\caption{\textbf{Application of local detunings.} \textbf{a, }In addition to the SLM used for atom trapping, a second SLM generates a set of superimposed tweezers that can apply arbitrary detuning profiles to the atoms. The pair of crossed AODs for local hyperfine control \cite{bluvstein2024logical} are not used in this work. \textbf{b,} A spin-echo sequence is used to measure and calibrate the differential AC Stark shift of the local detuning beams on the ground-state clock transition. Darker curves indicate a higher target detuning, weighted between 1 and 4, and the grey curves for unaddressed sites demonstrate negligible crosstalk between sites. The pattern here corresponds to a rotated version of the domains in the main text. \textbf{c, } Measured differential AC Stark shift of the  $5S_{1/2}\rightarrow5P_{3/2}$ transition. Unless stated otherwise (e.g., in Extended Data Fig.~\ref{fig:higgs_amplitude}), the  laser amplitude used is 1, corresponding to $\approx 160$\,$\mu$W per site and imparting an approximate light shift on the differential $\ket{g}\rightarrow\ket{r}$transition of $\delta_0 = -12(2)$MHz. The amplitude is the square root of power relative to $160$\,$\mu$W . When sites are weighted, the total power imparted on the array remains constant but is proportionally redistributed. \textbf{d, }The local detunings are used to prepare deterministic checkerboard orderings, both for a single global order (left) or for domains with different orders (right). A site polarization of unity, measured at the start of the hold time for varying $\Delta/\Omega$, corresponds to $\ket{g}$ ($\ket{r}$) for addressed (unaddressed) sites, while $-1$ corresponds to the flipped state. Note that at finite $\Delta/\Omega$, the polarization of $\ket{r}$ sites $\neq 1$; at large $\Delta/\Omega$, mean-field shifts reduce the spin polarization of both $\ket{g}$ and $\ket{r}$ along the domain wall. \textbf{e, } Dependence of the site polarization on $\Delta/\Omega$ in the bulk of the domain. As the local light shift becomes weaker relative to the global detuning, the probability of preparing $\ket{g}$ sites decreases slightly.
}
\label{fig:localcontrol_supp}
\end{figure*}

\begin{figure*}
\centering
\includegraphics[width=2\columnwidth]{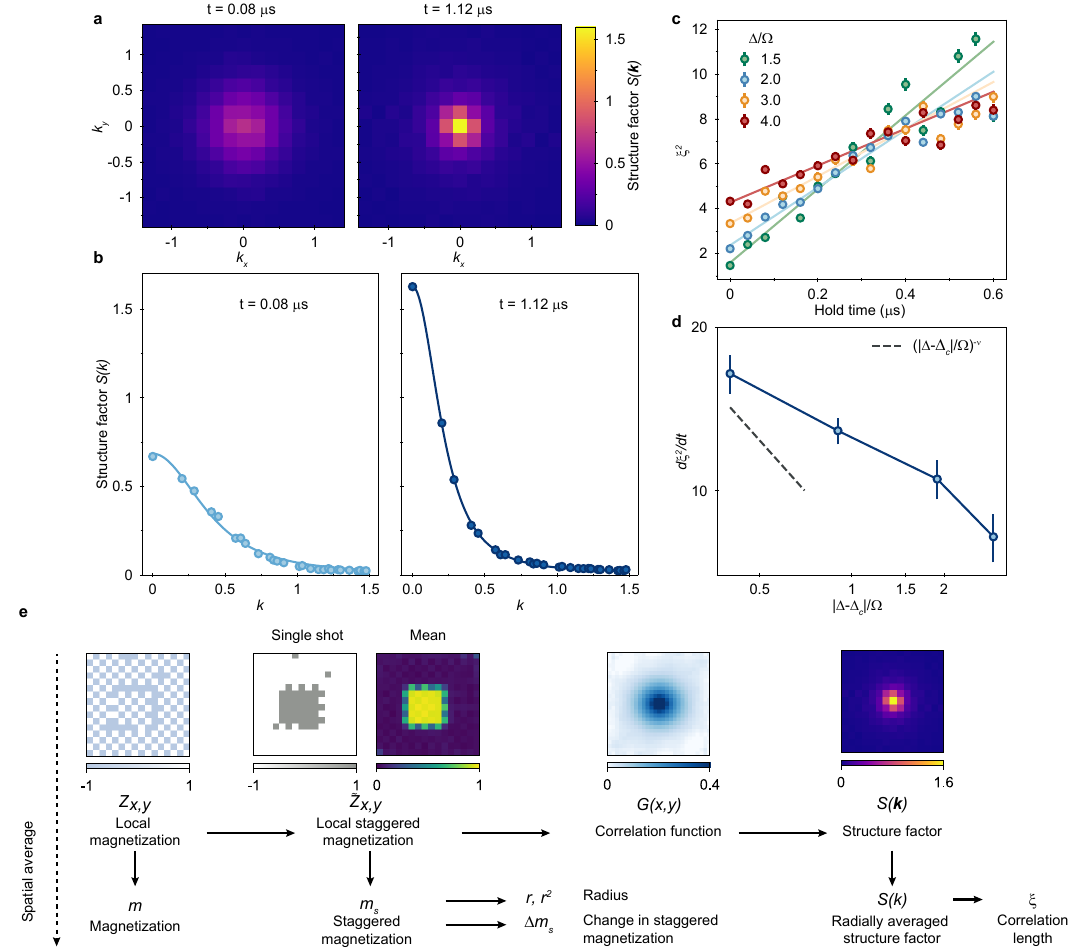} 
\caption{\textbf{Correlation lengths and structure factor.} \textbf{a, } Structure factor $S(\textbf{k})$ calculated at early (left) and late (right) times. \textbf{b, } Radially averaged structure factor $S(k)$ at early (left) and late (right) times for $\Delta/\Omega = 2.0$. The correlation length $\xi$ is extracted by fitting $S(k)$ to $b\xi^2/\pi(k^2\xi^2+1)^{3/2}$. \textbf{c, } For early times, the squared correlation length $\xi^2$ evolves linearly with time for all values of $\Delta/\Omega$ tested. \textbf{d, } Dependence of $\partial_t\xi^2$ on the distance from the quantum critical point. We find that the dynamics accelerate as  the system is held closer to the phase transition. The departure from the theoretically expected scaling (dashed line) stands in contrast to the case with locally seeded domains in Fig.~\ref{fig3:square defect}d and points to the possible role of finite-size effects due to boundaries for these global sweeps.
\textbf{e, } Summary of the relationship between main observables used throughout the manuscript. By imaging the atoms, we obtain the local magnetization of the system, from which the local staggered magnetization is calculated. Other key observables are then derived from the local staggered magnetization.}

\label{fig:corr_lengths}
\end{figure*}

\newpage

\begin{figure*}
\centering
\includegraphics[width=2\columnwidth]{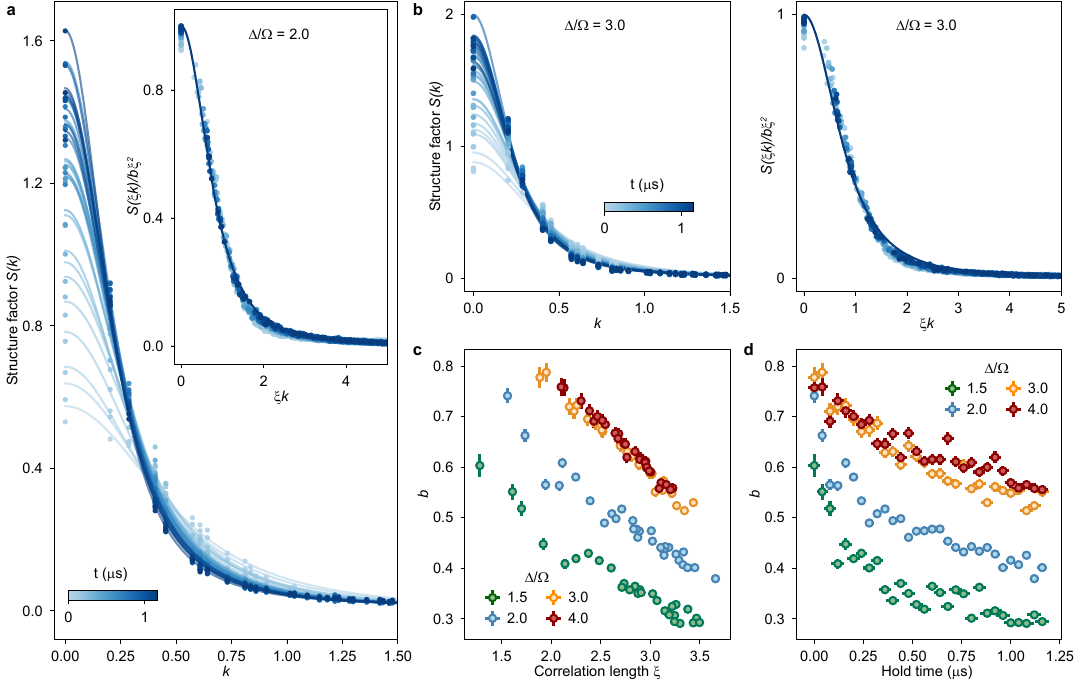} 
\caption{\textbf{Data collapse.} \textbf{a, } The structure factor $S(k,t)$ of the staggered magnetization, for different hold times, at $\Delta/\Omega=2.0$. Inset: the structure factors collapse onto a single curve when the time axis is rescaled by the correlation length $\xi(t)$ and the magnitude of the structure factor is rescaled by a time-dependent amplitude $\xi^2(t)b(t)$. \textbf{b, } At various values of $\Delta/\Omega$, the structure factors at different time (left) collapse onto a single curve (right). In this case, we show data for $\Delta/\Omega = 3.0$. \textbf{c, } We find a clear  dependence of amplitude $b$ on the nonequilibrium correlation length $\xi$, and that their relation is $\Delta$-dependent. \textbf{d, } We further plot the dependence of amplitude $b$ on hold time.}
\label{fig:collapse}
\end{figure*}

\newpage

\begin{figure*}
\centering
\includegraphics[width=2\columnwidth]{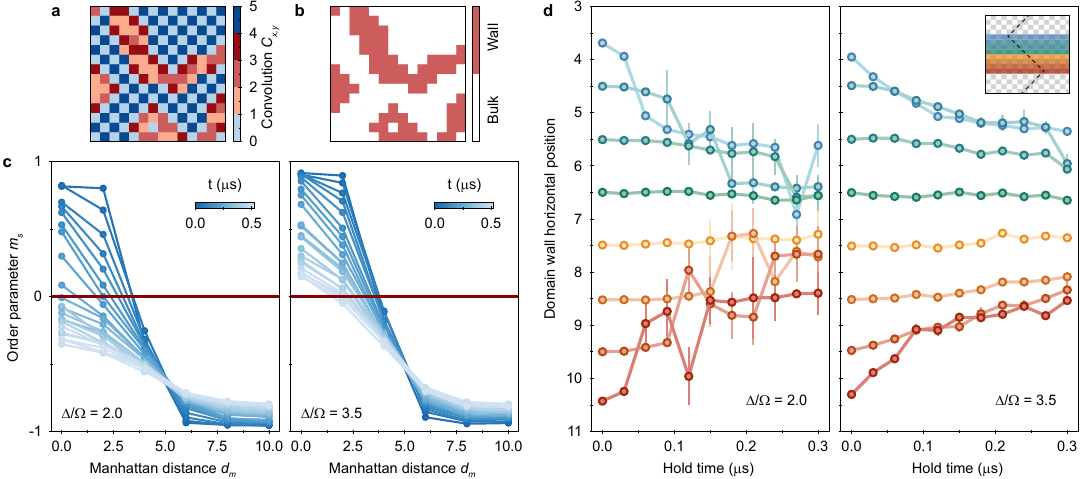}
\caption{\textbf{Analysis details on domain wall dynamics.} \textbf{a, } Convolution to calculate the coarse-grained-local staggered magnetization for each single shot. This is used to distinguish the bulk checkerboard orderings from the domain walls. Extremal values of 0 and 4 indicate the ideal coarse-grained local staggered magnetization of either order while values of 1--3 indicate deviations. \textbf{b, } Demarcation of domain wall boundaries and the bulk for the same shot as shown in \textbf{a}. Red (white) indicates the domain walls (bulk). \textbf{c, } Staggered magnetization radially averaged at each Manhattan distance from the center of the prepared square presented in Fig.~\ref{fig3:square defect}. The radius is extracted as the $d_m$ at which the linearly interpolated $m_s$ crosses 0 (red lines). \textbf{d, } Sweeps to various end detunings for the motion of the zigzag domain wall shown in Fig.~\ref{fig4:zigzag}c. The acceleration of the dynamics near the QPT is also clearly supported for this shape of the domain wall.  For both \textbf{c} and \textbf{d}, the left and right plots show data for $\Delta/\Omega = 2.0$ and $\Delta/\Omega = 3.5$, respectively. All data shown is for $\Omega/2\pi=6$\,MHz. 
}
\label{fig:local_analysis}
\end{figure*}

\newpage

\begin{figure*}
\centering
\includegraphics[width=2\columnwidth]{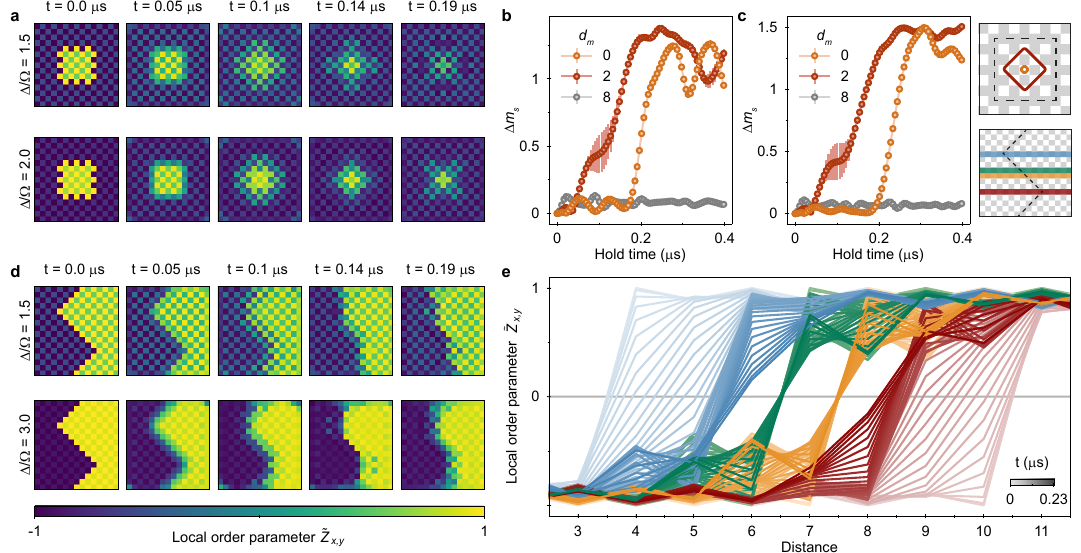}
\caption{\textbf{Numerical simulations of locally prepared domains.} \textbf{a, } TDVP simulations of coarsening dynamics observed when the system is initialized with a single square domain in the center (performed on a $15\times15$ lattice). The initial state approximates the experimental state preparation in Fig.~\ref{fig3:square defect}. \textbf{b, c, } The change in the radially averaged local staggered magnetization at Manhattan distances $d_m = 0,2,8$ from the center of the injected square. As observed in Fig.~\ref{fig3:square defect}e, the layer at $d_m =2$,  closer to the domain boundary, changes order before $d_m = 0$. We also find that similarly, the atoms in the bulk ($d_m=8$) remain in their initially prepared order. In these simulations, a slight acceleration in the dynamics is also observed on going closer to the phase transition. \textbf{d, } Simulation of the zigzag initial state described in Fig.~\ref{fig4:zigzag} on a $16\times 15$ lattice. \textbf{e, } As in the experimental results, the locally curved points of the domain wall move towards the center of the domain while points with no domain-wall curvature remain stable. Colors follow those also shown in the inset of Fig.~\ref{fig4:zigzag}. 
}
\label{fig:local_numerics}
\end{figure*}

\newpage

\begin{figure*}
\centering
\includegraphics[width=2\columnwidth]{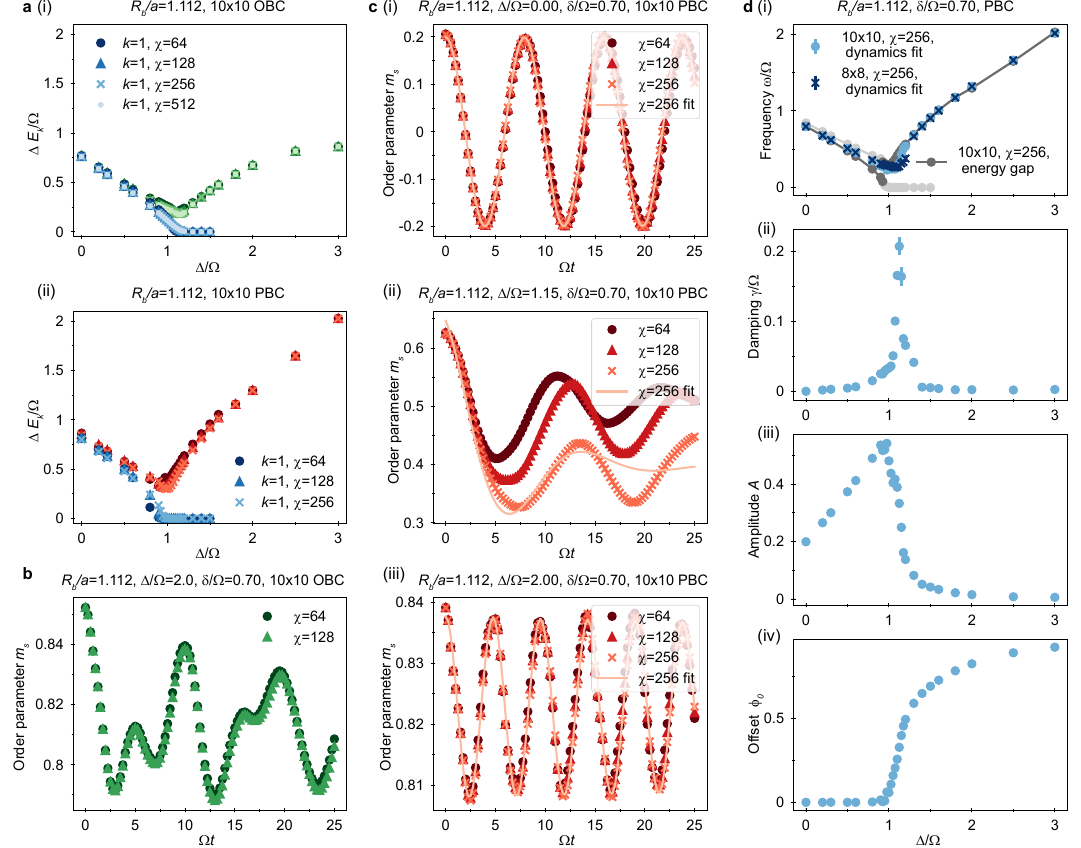}
\caption{\textbf{Numerical simulations of the amplitude mode.} \textbf{a,} Bond-dimension scaling of the energy gaps $\Delta E_1$, $\Delta E_2$ of the first two excited states above the ground state as a function of $\Delta/\Omega$, obtained via DMRG on a $10\times 10$ lattice with (i) open boundary conditions (OBC), and (ii) periodic boundary conditions (PBC). \textbf{b,} Order-parameter dynamics at $\Delta/\Omega=2$, of an initial state prepared with pinning fields $\delta_l/\Omega=-0.7$, on a $10\times 10$ lattice with OBC, simulated via TDVP with $\chi=64,128$.
\textbf{c, } Order-parameter dynamics at (i) $\Delta/\Omega=0$, (ii) $\Delta/\Omega=1.15$, (iii) $\Delta/\Omega=2$, of an initial state prepared with pinning fields $\delta_l/\Omega=-0.7$, on a $10\times 10$ lattice with PBC, simulated via TDVP with $\chi=64,128,256$. The order-parameter dynamics are well modeled as a damped harmonic oscillator, as shown by the fitting. \textbf{d, } Oscillation parameters (blue): (i) frequencies, (ii) damping rates, (iii) offsets, and (iv) amplitudes as a function of $\Delta/\Omega$, obtained from functional fits to the numerical simulations on a $10\times 10$ PBC lattice with $\chi=128$ shown in \textbf{c}. The frequencies agree with the bulk (PBC) energy gaps for a $10\times 10$ PBC lattice obtained in panel \textbf{a}(ii), and with the oscillation frequencies for a $8\times 8$ PBC lattice, for comparison. 
}
\label{fig:higgs_supp}
\end{figure*}
\newpage

\begin{figure*}
\centering
\includegraphics[width=2\columnwidth]{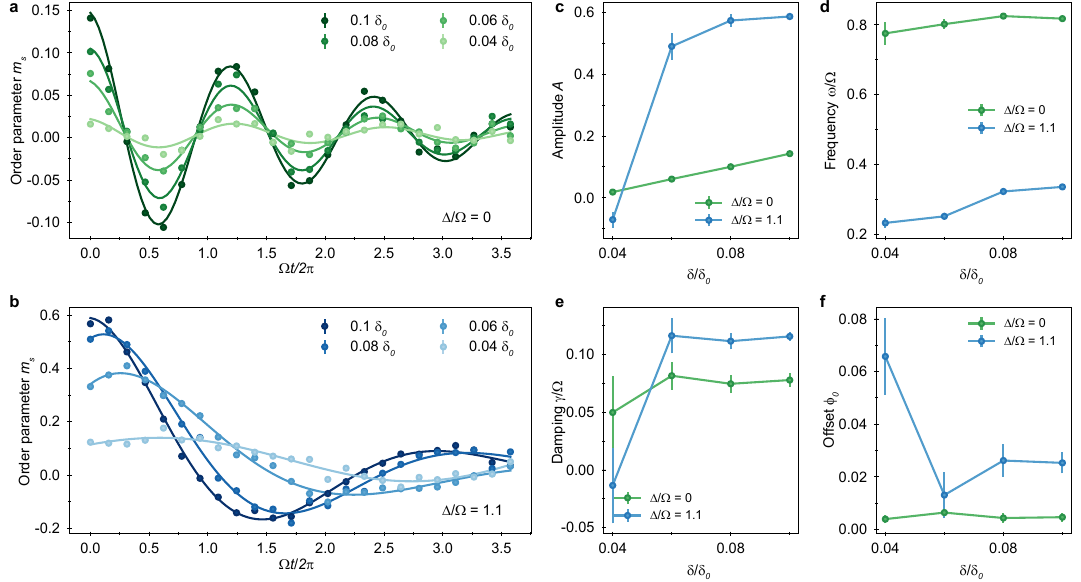}
\caption{\textbf{Effect of applied local detunings on the Amplitude mode} \textbf{a, } Amplitude mode oscillations in the disordered phase at $\Delta/\Omega = 0$. The magnitude of the applied local detunings is varied as a fraction of $\delta_0 = -2\pi\times 12(2)$MHz $\approx-4\Omega$ (see Extended Data Fig.~\ref{fig:localcontrol_supp}c). \textbf{b, } ``Higgs''-mode oscillations for varying magnitudes of the local detunings at $\Delta/\Omega = 1.1$. Note that for \textbf{a} and \textbf{b}, the frequency extracted here for $\delta = 0.06 \delta_0 \approx -\Omega/4 $ is plotted in yellow for comparison in Fig.~\ref{fig5}. \textbf{c-d, } With decreasing magnitude of $\delta$, we observe a decrease in the oscillation amplitude (\textbf{c}) and frequency (\textbf{d}). The dependence is more pronounced closer to the phase transition. This behaviour can be qualitatively captured by considering the anharmonicity of the Landau mean-field potential in Eq.~\eqref{eq:lagrangian} as the phase transition is approached. \textbf{e, } As also seen in Fig.~\ref{fig5}, the damping is stronger near the phase transition for both detunings. \textbf{f, } The oscillations at $\Delta/\Omega = 1.1$ are centered around finite $\phi_0$ suggesting a transition into the ordered phase at this value. The data shown here is taken for the same $\Omega$ as all other ``Higgs''-mode oscillations, $\Omega/2\pi = 3.1$\,MHz.} 
\label{fig:higgs_amplitude}
\end{figure*}

\newpage

\begin{figure*}
\centering
\includegraphics[width=2\columnwidth]{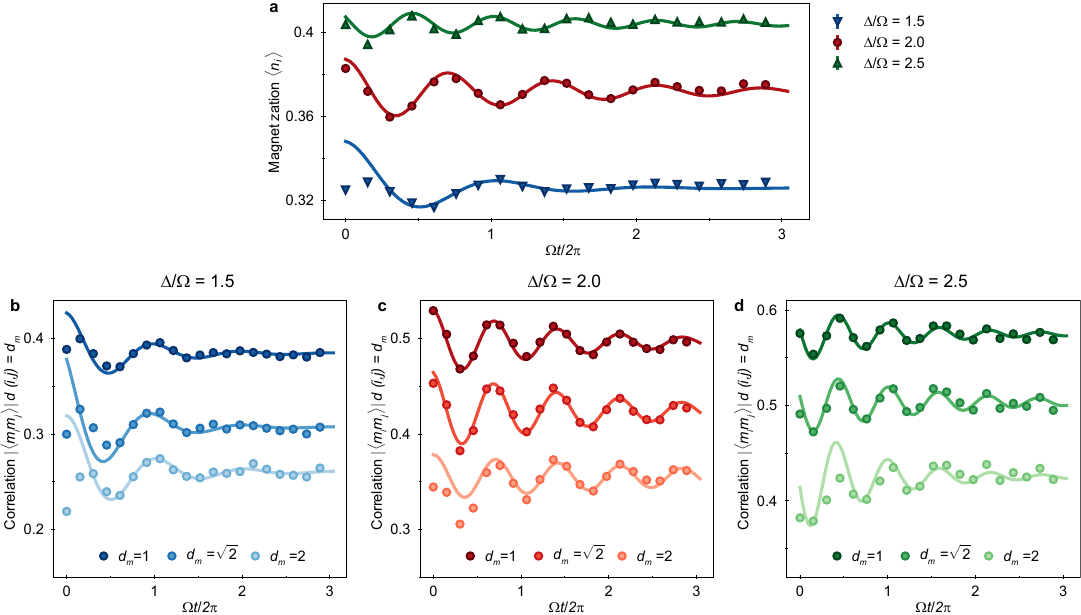}
\caption{\textbf{Amplitude mode after a sweep through the QPT.} ``Higgs''-mode oscillations are apparent in parallel to coarsening in multiple observables. The oscillations presented here are observed following the protocol described in Fig.~\ref{fig:correlations}, with $\Omega/2\pi = 3.8$\,MHz. Note that we fit the the oscillation after an initial decay on the time-scales less than a Rabi-cycle. Extracted frequencies are consistent across observables. \textbf{a, } Oscillations of the magnetization $\langle n_i\rangle$. \textbf{b, c, d,} Dynamics of the two-point correlation function in position space for various end detunings. We find that the two-point correlators exhibit oscillations at frequencies for various values of $\Delta/\Omega$ that are consistent with the ground-state energy gap shown in Fig.~\ref{fig5}. 
}
\label{fig:higgs_global}
\end{figure*}

\newpage

\begin{figure*}
\centering
\includegraphics[width=2\columnwidth]{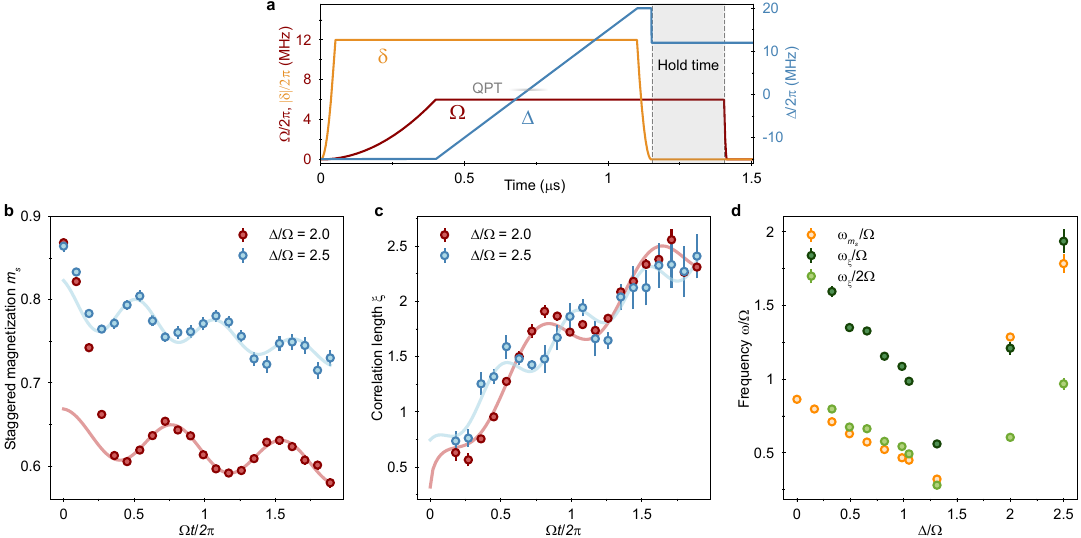}
\caption{\textbf{Ordered-phase quenches and frequency doubling.} \textbf{a,} Sweep profile for quenches into the ordered phase. Under a constant Rabi dive at $\Omega/2\pi = 6$\,MHz, the global detuning $\Delta$ is ramped from large negative values to large positive values deep into the ordered phase ($\Delta/\Omega = 3.3)$. The local detuning $\delta$ is quenched off quickly over 50\,ns. The global $\Delta$ is then also quenched down to varying final detuning values closer to the QPT. 
\textbf{b, } Oscillations of the staggered magnetization in the ordered phase following this preparation sequence. In agreement with the Landau mean-field picture, these oscillations occur around a nonzero value of the order parameter $m_s$. Note that we fit the the oscillation after an initial decay on the time-scale less than a Rabi-cycle.
\textbf{c, } Oscillations in the correlation length on top of a growing background. 
\textbf{d, } Ratio between oscillation frequencies extracted from the staggered magnetization ($\omega_{m_s}$) and the correlation length ($\omega_{\xi}$). In the disordered phase, $\omega_{\xi}/\omega_{m_s} \approx 2$ while in the ordered phase, this changes to  $\omega_{\xi}/\omega_{m_s} \approx 1$. 
}
\label{fig:ordered_quenches}
\end{figure*}


\begin{figure*}
\centering
\includegraphics[width=2\columnwidth]{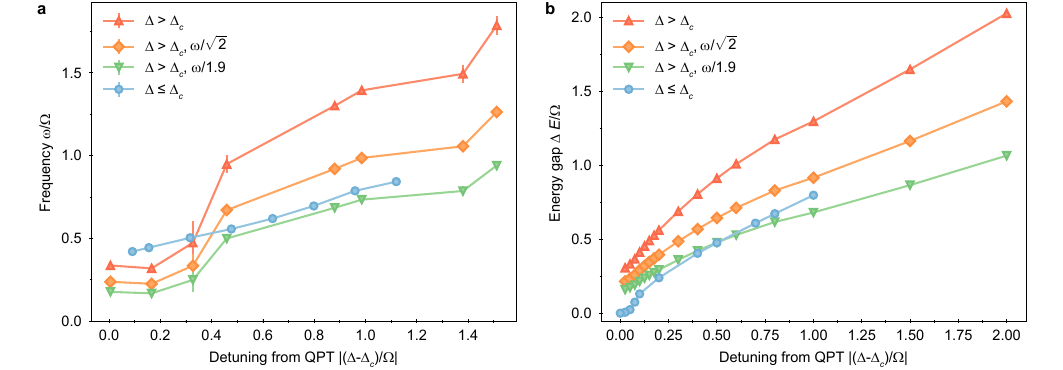} 
\caption{\textbf{Ratio between oscillation frequencies.} \textbf{a, }Experimentally extracted oscillation frequency of the amplitude mode for the full data set shown in Fig.~\ref{fig5}a of the main text. The frequency is plotted with respect to the distance from the critical point, $\Delta_c = 1.12$ measured in \cite{ebadi2021quantum}. We see closer agreement with the theoretically predicted factor, $\omega(-|q|)/\omega(|q|) = 1.9$ \cite{sachdev1997theory, brezin1974, caselle1997, campostrini2002} than with the Landau mean-field ratio of $\sqrt{2}$. However, deviations from both factors close to the phase transition are apparent. \textbf{b, } Numerical simulations using MPS methods on 10x10 sites with periodic boundary conditions described in Extended Data Fig.~\ref{fig:higgs_supp}. 
We again see approximate agreement with the expected ratio, as well as deviations near the numerically determined critical point $\Delta_c=1.0$. 
}
\label{fig:higgs_ratio}
\end{figure*}

\clearpage

\end{document}